\newcommand{\diracslash}[1]{#1\llap{/\kern2pt}}

\def\bearr{\begin{eqnarray}}
\def\eearr{\end{eqnarray}}

\newcommand{\be}{\begin{equation}}
\newcommand{\ee}{\end{equation}}
\newcommand{\bea}{\begin{eqnarray}}
\newcommand{\eea}{\end{eqnarray}}
\newcommand{\ba}[1]{\begin{array}{#1}}
\newcommand{\ea}{\end{array}}

\newcommand{\eqrf}[1]{Eq.\ (\ref{#1})}

\newcommand{\eqrftw}[2]{Eqs.\ (\ref{#1}) and (\ref{#2})}
\newcommand{\eqrftr}[3]{Eqs.\ (\ref{#1},\ref{#2},\ref{#3})}

\documentclass[preprint,secnumarabic,floats,amssymb,nobibnotes,nofootinbib,aps,prc,showpacs,tightenlines]{revtex4}
\usepackage{graphicx}
\usepackage{amsmath}
\usepackage{latexsym}
\usepackage{epsfig}
\usepackage{subfigure}
\usepackage{multirow}
\begin{document}

\title{Reheating constraints on K-inflation}
\author{Pooja Pareek and Akhilesh Nautiyal}
\affiliation{Department of Physics, Malaviya National Institute of Technology, Jaipur, 
JLN  Marg, Jaipur-302017, India}
\begin{abstract}
In this work we revisit constraints 
on K-inflation with DBI kinetic term and 
power-law kinetic term from reheating. For DBI kinetic term we choose monomial potentials,
$V\propto \phi^n$ with $n=2/3\,,1\,,\,2$ and $4$, 
and natural inflaton potential, and for power-law kinetic term we choose quadratic, quartic and
exponential potentials. The phase of reheating can be parameterized in terms of
reheating temperature $T_{re}$, number of e-folds during reheating $N_{re}$ and 
effective equation of state during reheating $w_{re}$. These parameters can be  related  to the 
spectral index $n_s$ and other inflationary parameters depending on  the choice of inflaton 
kinetic term and potential. 
By demanding that $w_{re}$ should have a finite range and $T_{re}$ should be above
electroweak scale, one can obtain the bounds on $n_s$ that can provide bounds on tensor-to-scalar
ratio $r$. We find, for K-inflation with DBI kinetic term and quadratic and quartic potentials,
that the upper bound on $r$ for physically plausible value of $0\le w_{re} \le 0.25$ is 
slightly larger than the Planck-2018 and BICEP2/Keck array bound, and for $n=2/3$ and $1$, 
the reheating equation of state should be less than $0$ to satisfy Planck-2018 joint  
constraints on $n_s$ and $r$. However, natural inflation with
DBI kinetic term is compatible with Planck-2018 bounds  on $r$ and joint constraints on 
$n_s$ and $r$ for physically plausible range $0 \le w_{re}\le 0.25$.
The quadratic and quartic potential with power-law kinetic term are also compatible with 
Planck-2018 joint constraints on $n_s$ and $r$ for $0\le w_{re} \le 1$. However, for 
exponential potential with power-law kinetic term, the equation of state during reheating 
$w_{re}$ should be greater than $1$ for $r-n_s$ predictions to lie within
$68\%$C.L. of  joint constraints on $n_s$ and $r$ from Planck-2018 observations.  
\end{abstract}


\maketitle

\section{Introduction}

 The idea of inflation \cite{Guth:1980zm} is now well accepted solution to the 
horizon and flatness problem of big-bang cosmology. It  also provides seeds for 
anisotropy of cosmic microwave background and structures in the 
universe \cite{Mukhanov:1981xt,Starobinsky:1982ee,Guth:1985ya}.
The predictions of inflation, i.e, nearly scale-invariant, Gaussian and adiabatic density
perturbations are confirmed by the various CMB  observations such as
COBE \cite{Smoot:1992td} WMAP \cite{Komatsu:2010fb}, Planck \cite{Akrami:2018odb} etc.
In standard scenario potential energy of a scalar field, named as inflaton, dominates the 
energy density of the universe during inflation and provides quasi exponential expansion.
Inflaton rolls slowly through its potential during inflation, and the quantum fluctuations in 
this field, which are coupled to the metric fluctuations, generate the primordial density 
perturbations (scalar perturbations). The vacuum fluctuations in the tensorial part of the 
metric generated during inflation are responsible for the primordial gravitational waves 
(tensor perturbations). The power spectra for scalar and tensor perturbations generated during
inflation depend on the inflaton potential, which can be obtained from particle physics models
and string theory. Many models of inflation have been explored in recent 
years (see \cite{Martin:2013tda} for details). 
Although the predictions of  inflation are in excellent agreement with the CMB observations,
 we still lack a unique model. 
The most popular quadratic and quartic potentials are ruled out by 
recent Planck observations \cite{Akrami:2018odb} as they give large tensor-to-scalar ratio.

There is an  alternative to the standard scenario of inflation, named as $K-$inflation
\cite{ArmendarizPicon:1999rj,Garriga:1999vw}, 
where inflation is achieved by the non-standard kinetic term of the inflaton. The 
non-standard kinetic term in the action of inflaton can have monomial and polynomial form
\cite{ArmendarizPicon:1999rj,Mukhanov:2005bu}  or 
 Dirac-Born-Infield  form \cite{Gibbons:2002md}, which arises 
 in string theory \cite{Sen:1999md,Gibbons:2000hf,Sen:2000kd,Sen:2002nu}
 (see \cite{Piao:2002vf,Mazumdar:2001mm,Chingangbam:2004ng,Choudhury:2015hvr} for 
various choices of noncannonical kinetic terms and potentials derived from string theory). 
In \cite{Devi:2011qm,Li:2012vta,Unnikrishnan:2012zu} it was
shown that the tensor-to-scalar ratio can be lowered for quadratic and quartic potentials 
with noncanonical kinetic term.  K-inflation with pseudo-Nambu-Goldstone-Boson has also been
studied in \cite{Devi:2011qm,Rashidi:2017chf,Bhattacharya:2018xlw} and it is shown that
 natural inflation
with noncanonical kinetic term is compatible with the Planck CMB observations. 
Power-law kinetic
term with exponential potential has also been studied in \cite{Unnikrishnan:2012zu} and it is
found that this model is also compatible with the CMB observations. 
In \cite{Lola:2020lvk} power-law kinetic term has also been studied with deformed steepness
exponential potentials.   

Several generalizations of
K-inflation have been studied in the literature such as inflaton with non-minimal 
coupling with Ricci scalar \cite{Oikonomou:2021edm,Sen:2008bg}, 
inflaton coupled with Gauss-Bonnet invariant \cite{Odintsov:2021lum} and K-inflation with
$f(R)$ gravity \cite{Nojiri:2019dqc}. K-inflation with constant-roll conditions has also been
 studied in \cite{Odintsov:2019ahz}. It has been shown in \cite{Gialamas:2019nly} that the 
action of $R^2$-inflation in the framework of Palatini gravity resembles K-inflation 
models in the Einstein frame. 

All these models of noncanonical inflation are in agreement with the current bounds on spectral 
index and tensor-to-scalar ratio from Planck-2018 observations, and there is no unique choice
for noncanonical kinetic term and inflaton potential. 

 At the end of inflation, the universe reaches to a 
cold and highly non-thermal state without any matter content.
However, for baryogenesis and  big-bang nucleosynthesis the universe needs to be in a 
thermalized state at a very high temperature. This is achieved by reheating, a transition phase between the end of inflation
and start of radiation dominated era. During this phase the inflaton energy is transferred to 
radiation, baryons and leptons, leaving the universe at a reheating temperature $T_{re}$ at the
onset of radiation epoch. In the simplest models of reheating 
\cite{Abbott:1982hn,Dolgov:1982th,Albrecht:1982mp} inflaton oscillates around the 
minimum of its potential and   decays perturbatively into the standard model particles
through various interactions of inflaton with other scalars and fermions.
However, perturbative reheating is model dependent and cannot give correct description of 
the process at various states, and it also does not take into account the 
coherent nature of the inflaton field \cite{Traschen:1990sw,Dolgov:1989us}. In other
scenarios the reheating is preceded by preheating, during which the classical inflaton field 
decays into massive particles via non-perturbative 
processes such as parametric resonance \cite{Kofman:1994rk,Kofman:1997yn}, 
tachyonic instability \cite{Greene:1997ge,Dufaux:2006ee}, and instant preheating 
\cite{Felder:1998vq}. 
After preheating these massive particles decay perturbatively into the 
standard model particles, which are then thermalized and the universe enters into 
radiation dominated era with a blackbody spectrum at a temperature $T_{re}$, named as
reheating temperature. 

Although the physical processes involved during reheating are complex, this phase
can be parameterized in terms of three parameters, reheating temperature $T_{re}$, the effective
equation of state of matter during reheating $w_{re}$ and duration of reheating that is
 given in terms of number of e-foldings $N_{re}$.
The reheating temperature cannot be constrained from CMB and LSS observations,
but, it is assumed that $T_{re}$ should be above the electroweak scale so that the weak scale
dark matter can be produced. In a more conservative approach $T_{re}$ should be above $10$ Mev
for successful big-bang nucleosynthesis. The reheating temperature can be as low as 
$2.5$ to $4$ MeV, for considering late-time entropy production by massive particle decay 
\cite{Kawasaki:1999na,Kawasaki:2000en}. By considering instant reheating we can also 
put an upper bound on the reheating temperature $T_{re}$ to be of the order of scale of 
inflation, which is $10^{16}$ GeV for current upper bounds on tensor-to-scalar ratio from Planck.
The second parameter of reheating is effective equation of state $w_{re}$ representing evolution 
of energy density of the cosmic fluid during reheating. This parameter is, in general, 
time dependent and its value changes from $-\frac13$ to $\frac13$ from the end of inflation
to the onset of radiation dominated era. For the reheating occurring due to perturbative decay of 
massive inflaton, $w_{re}$ is $0$ and for instant reheating it is $\frac13$.
The evolution of equation of state during preheating and the early thermalization state
was studied in \cite{Podolsky:2005bw} by using lattice numerical simulation for quadratic
potential  interacting with light fields, and it was found that  the equation of state
starts from  $w_{re}=0$ after inflation and saturates around $w_{re}\sim 0.2-0.3$ long
before the thermalization of the universe. This analysis was generalized in 
\cite{Lozanov:2016hid,Lozanov:2017hjm} for inflaton potentials 
behaving as $|\phi|^{2n}$ near $|\phi|=0$ and flatter beyond some scale $|\phi|=M$ by taking into
account the fragmentation of the inflaton field and ignoring coupling to massless fields, and it
was found that the equation of state $w_{re}$ reaches  $1/3$ for $n>1$ after sufficient 
long time, while, it remains $0$ for $n=1$.
 The third parameter to describe reheating is its duration, which can be defined in terms
of number of e-foldings from the end of inflation to the beginning of radiation dominated 
epoch. This duration is incorporated in the number of e-foldings $N_k$ during inflation from 
the time, when the Fourier mode $k$ corresponding to the horizon size of  present 
observable universe leaves the Hubble radius during inflation, to the end of inflation. The 
e-foldings $N_k$ depends on the potential of inflaton and it should be between $46$ to $70$
to solve horizon problem. The upper bound on $N_k$ arises from assuming that the universe
reheats  instantaneously, and  the lower bound comes from considering the reheating temperature
at the electroweak scale. In \cite{Liddle:2003as,Dodelson:2003vq} a detailed analysis of upper 
bound on $N_k$ we performed for
various scenarios and it was shown that, for some cases, $N_k$ can be as large as $107$.

In \cite{Dai:2014jja,Munoz:2014eqa,Cook:2015vqa} it was shown that 
the above mentioned reheating parameterization can be used to constrain various models of 
inflation. The reheating temperature $T_{re}$ and the e-folds during reheating 
$N_{re}$ can be expressed in terms of spectral index $n_s$ by assuming $w_{re}$ to 
be constant during reheating \cite{Dai:2014jja,Munoz:2014eqa,Cook:2015vqa}.  
By imposing that the effective equation of state during 
reheating lies between $0$ and $0.25$ and the temperature at the end of 
reheating $T>100$ GeV, one can obtain bounds on spectral index $n_s$ and $N_k$, which 
translates to bounds on tensor-to-scalar ratio. As various models of inflation predict
similar values of $n_s$ and $r$, it has been shown in \cite{Mishra:2021wkm} that by imposing 
constraints on these reheating parameters this degeneracy can be removed.
The bounds on reheating parameters were also used to constrain 
tachyon inflation \cite{Nautiyal:2018lyq}, where inflaton have a DBI kinetic term with 
inverse $\cosh$ and exponential potential. 
It was shown that one requires effective equation of state during reheating $w_{re}>1$ to 
satisfy Planck-2018 observations.  

In this work we use these reheating parameters to constrain K-inflation with
DBI kinetic term with monomial potentials and PNGB potential, and K-inflation with 
power-law kinetic term with monomial and exponential potentials. Reheating constraints on 
noncanonical inflation with inflaton having DBI kinetic term and PNGB potential are already 
considered in \cite{Rashidi:2017chf} with Planck-2015 data. Here we revisit tachyon natural 
inflation with Planck-2018 data along with other potentials with DBI kinetic term. 

The work is organized as as follows: in section \ref{kigf} we discuss the dynamics of K-inflation
and present expressions for power spectra. 
In section~\ref{reheatparam} we discuss parameterization
of reheating phase. We obtain expressions for $T_{re}$ and $N_{re}$ in terms of spectral 
index by assuming constant effective equation of state during reheating. 
In section~\ref{dbikinetic} we discuss noncanonical inflation with DBI kinetic term and 
obtain expressions for $T_{re}$ and $N_{re}$ for monomial and PNGB potential for various
choices for $w_{re}$. We use these three parameters to constrain K-inflation with DBI kinetic 
term. In section~\ref{powerlawkinetic} we discuss dynamics of noncanonical inflation
with power law kinetic term, and obtain $T_{re}$ and $N_{re}$ for monomial and exponential 
potential with various choices of $w_{re}$. We again use these three parameters to constrain
K-inflation with power-law kinetic term.  In section~\ref{conclusion}
we conclude our work. 

\section{K-inflation: General framework} \label{kigf}
In K-inflation the inflaton field has a noncanonical kinetic term. The action for inflaton
is given as
\be
S = \int\sqrt{-g}\left\{-\frac{1}{16\pi G}R + {\cal L}\left(X,\phi\right)\right\}, 
\label{genaction}
\ee 
where ${\cal L}\left(X,\phi\right)$ is the Lagrangian of scalar field, which is a function of
kinetic term $X=\frac12\partial_\mu\phi\partial^\mu\phi$ and the field $\phi$.
We can obtain energy-momentum tensor by  varying this action with respect to the metric as
\be
T_{\mu\nu} = \frac{\partial{\cal L}\left(X,\phi\right)}{\partial X} 
\partial_\mu\phi\partial_\nu\phi-{\cal L}\left(X,\phi\right)g_{\mu\nu}.
\label{tmunu}
\ee
This energy-momentum tensor is equivalent to  that of a  perfect fluid with pressure
\be
p = {\cal L}\left(X,\phi\right), \label{pgen}
\ee
energy density 
\be
\rho = 2X\frac{\partial{\cal L}}{\partial X}-{\cal L}\label{rhogen}
\ee
and four-velocity
\be
u_\mu = \sigma\frac{\partial_\mu\phi}{\sqrt{2X}}, \label{umugen}
\ee
where $\sigma$ refers to the sign of $\dot\phi$.
The evolution of the universe is described using Friedmann equations
\bea
&&H^2 = \frac{1}{3M_P^2}\rho, \label{hsquare}\\
&&\dot H = - \frac{1}{2M_P^2} \left(\rho+p\right). \label{hdot}
\eea
Here $M_P=\frac{1}{\sqrt{8\pi G}}$ is the reduced Planck mass. 
For inflation the second derivative of the scale factor should satisfy the condition 
$\frac{\ddot a}{a} = \dot H + H^2>0$, which can be expressed in terms of the slow-roll parameter
\be
\epsilon = -\frac{\dot H}{H^2}<1.
\ee
For our analysis we define slow-roll parameters in terms of  the Hubble flow parameters as
\cite{Schwarz:2001vv} 
\begin{align}
\epsilon_{0}\equiv\frac{H_k}{H}\label{eps0},
\end{align}
and
\begin{align}
\epsilon_{i}\equiv\frac{d\ln|\epsilon_i|}{dN},\; \; i\geq{0} \label{epsi}.
\end{align}
where $H_k$ is the Hubble constant during inflation at the time when a particular mode
$k$ leaves the horizon and $N$ is the number of e-foldings
\be
N = \ln\left(\frac{a}{a_i}\right),\label{Naai}
\ee
where $a_i$ is the scale factor at the beginning of inflation. The first derivative of 
Hubble flow parameter with respect to time can be expressed as
\be
\dot\epsilon_i = \epsilon_i\epsilon_{i+1}.
\ee
The first two Hubble flow parameters $\epsilon_1$ and $\epsilon_2$ can be obtained in 
terms of energy density and pressure as
\bea
&&\epsilon_1=\epsilon = \frac{3}{2}\frac{\rho+p}{\rho}\label{ep1gen}\\
\text{and}\;\; &&\epsilon_2 = \frac{3}{2H}\frac{d}{dt}\left(\frac{\rho+p}{\rho}\right)
\label{ep2gen}.
\eea
The power spectra for scalar and tensor perturbations, scalar spectral index $n_s$ and 
tensor-to-scalar ratio $r$ for K-inflation are computed in \cite{Garriga:1999vw}, 
and can be expressed in terms of the Hubble flow parameters as
\bea
P_{\zeta}&=&\frac{H^2}{8\pi^2M_{P}^2c_s\epsilon}\bigg{|}_{c_S k=aH},\label{scalar}\\
P_h&=&\frac{2}{\pi^2}\frac{H^2}{M_{P}^2}\bigg{|}_{c_S k=aH},\label{tensor}\\
n_s&=&1-2\epsilon_1-\epsilon_2,\label{nsk}\\
r&=&16c_S\epsilon_1\label{rttsr}.\\\nonumber
\eea
where
\be
 c_S^2=\frac{\partial p/\partial X}{\partial \rho/\partial X} \label{cs2} 
\ee
is the sound speed for perturbations. These power spectra are
evaluated at the Hubble crossing during inflation
 for the Fourier mode $k$ of curvature perturbation and tensor perturbation. In K-inflation
the condition for Hubble exit is modified as $c_S k = aH$ for scalar perturbations. For CMB 
analysis the power spectrum for curvature perturbation is expressed as 
$P_\zeta=A_S \left(\frac{k}{k_0}\right)^{n_s-1}$, where the amplitude of scalar perturbations
$A_S$ is given by \eqrf{scalar}. All the three quantities $A_S$, $n_s$ and $r$ are evaluated
at pivot scale $k_0$, which is $0.05$ M\textsubscript{pc}\textsuperscript{-1} for Planck 
observations, and they depend on the choice of noncanonical kinetic term and potential
of inflaton. Bounds on these quantities are provided by CMB and LSS observations, 
which can be used to  put constraints on parameters of the  potential and  the
noncanonical kinetic term of inflaton. 
Again all these inflationary parameters also appear in reheating temperature and number of 
e-folds during reheating, which can, along with CMB constraints, be used to analyze models of 
inflation. In this work we analyze K-inflation having noncanonical kinetic term of DBI form 
in section and of power-law form. In the next section  we obtain relation between 
reheating parameters, $T_{re}$ and $N_{re}$, and inflationary parameters. 

\section{Parameterizing reheating} \label{reheatparam}
As mentioned earlier the reheating phase can be parameterized in terms of thermalization 
temperature $T_{re}$ at the onset of radiation dominated epoch after reheating, effective
equation of state of cosmic fluid $w_{re}$ during reheating and number of e-folds $N_{re}$ for which reheating lasts. In our analysis we consider $w_{re}$ to be constant during reheating.
Its value should lie between $-\frac13$ to $1$. The lower bound on $w_{re}$ comes from the 
fact that it should be $-\frac13$ when inflation ends, and the upper bound arises from the 
fact that it should be smaller than $1$ to satisfy dominant energy condition of general 
relativity, $\rho \ge |p|$ for the causality condition to be preserved 
\cite{Munoz:2014eqa,Martin:2010kz,Chavanis:2014lra}.\\
In this section  we express the reheating parameters ($N_{re}$, $T_{re}$ and $w_{re}$) in 
terms of the quantities that are derivable from inflation 
models \cite{Martin:2014nya,Dai:2014jja,Mielczarek:2010ag,Easther:2011yq}. 
Assuming a constant equation of state during reheating and 
using $\rho\propto a^{-3\left(1+w\right)}$, the reheating epoch can be expressed as
\begin{equation}
\frac{\rho_{end}}{\rho_{re}}=\left( \frac{a_{end}}{a_{re}}\right)^{-3(1+w_{re})},\label{rhoendre}
\end{equation}
here  the subscript "${end}$" refers to the quantity evaluated at  the end of inflation, 
and the subscript "${re}$" denotes the quantity evaluated at  the end of reheating. 
  The number of e-foldings during reheating
is obtained using (\ref{rhoendre}) as
\bea
N_{re} &=& \ln \left(\frac{a_{re}}{a_{end}}\right) = \frac{1}{3(1+w_{re})} \ln\left(\frac{\rho_{end}}{\rho_{re}}\right)\nonumber\\.
 &=& \frac{1}{3(1+w_{re})} \ln\left(\frac{3}{2}\frac{V_{end}}{\rho_{re}}\right)\label{nrev}.
\eea
where we have used $\rho_{end}={\frac{3}{2}}V_{end}$ in the last expression as 
$ w =-\frac{1}{3}$ at the end of inflation.
At the end of reheating the universe enters into radiation era, hence
the energy density at the end of reheating can be expressed in terms of reheating temperature as
\begin{equation}
\rho_{re}= \frac{\pi^2}{30}g_{re} T_{re}^4\label{rhot},
\end{equation}
where $g_{re}$ is the number of relativistic species at the end of reheating. 
We will use $g_{re}=100$ (the value for standard model of particle physics) for our analysis.
Using \eqrftw{nrev}{rhot} $N_{re}$ can be expressed in terms of reheating temperature as
\begin{equation}
N_{re} = \frac{1}{3(1+w_{re})} \ln\left(\frac{30.\frac{3}{2}V_{end}}{\pi^2 g_{re}T_{re}^4}\right)\label{nrevt}.
\end{equation}

Since the entropy remains  conserved between the end of reheating and today, 
the reheating temperature can be related to  the  CMB temperature today as
\begin{equation}
T_{re} = T_0 \left(\frac{a_0}{a_{eq}}\right)\left(\frac{43}{11g_{re}}\right)^{1/3} = T_0 \left(\frac{a_0}{a_{eq}}\right) e^{N_{RD}}\left(\frac{43}{11g_{re}}\right)^{1/3}\label{trea},
\end{equation}
where ``$0$'' in the subscript denotes the values of the quantities evaluated at present epoch,
and ``$eq$'' refers to the values evaluated at matter-radiation equality.
 $N_{RD}$ in \eqrf{trea} refers to  the   number of e-foldings  during  radiation era, 
$ e^{-N_{RD}}\equiv \frac{a_{re}}{a_{eq}}$. The ratio $\frac{a_0}{a_{eq}}$ is expressed as
\begin{equation}
\frac{a_0}{a_{eq}} = \frac{a_0}{a_k}\frac{a_k}{a_{end}}\frac{a_{end}}{a_{re}}
\frac{a_{re}}{a_{eq}}=\frac{a_0 H_k}{c_S k} e^{-N_k} e^{N_{re}} e^{-N_{RD}}.\label{a0aeq}
\end{equation}
Here the subscript "$k$" denotes that the quantity is evaluated at the time  when Fourier 
mode $k$ crosses the Hubble radius during inflation. $N_k$ represents the number of e-folds
from this time  to the end of inflation, and the condition for
horizon crossing  $c_S k=a_k H_k$ is also used. 
Substituting \eqrf{a0aeq} into \eqrf{trea}, we obtain
\begin{equation}
T_{re} = \left(\frac{43}{11g_{re}}\right)^{1/3} \left(\frac{a_0 T_0}{c_S k}\right) H_k e^{-N_k} e^{-N_{re}}.\label{trenk}
\end{equation}
Again substituting \eqrf{trenk} into  \eqrf{nrevt}, one can find
\begin{equation}
N_{re} = \frac{4}{3(1+w_{re})}\left[\frac{1}{4} \ln\left(\frac{3^2.5}{\pi^2 g_{re}}\right) + \ln\left(\frac{V_{end}^{1/4}}{H_k}\right)+\frac{1}{3} \ln\left(\frac{11g_{re}}{43}\right) + \ln\left(\frac{c_S k}{a_0 T_0}\right) + N_k +N_{re}\right]\label{nrenk}.
\end{equation}
This, on solving for $N_{re}$, with assumption $w_{re}\neq \frac{1}{3}$, gives
\begin{equation}
N_{re} = \frac{4}{(1-3w_{re})}\left[\frac{-1}{4} \ln\left(\frac{3^2.5}{\pi^2 g_{re}}\right)-\frac{1}{3} \ln\left(\frac{11g_{re}}{43}\right) - \ln\left(\frac{c_S k}{a_0 T_0}\right)  - \ln\left(\frac{V_{end}^{1/4}}{H_k}\right)- N_k \right].\label{nre1}
\end{equation}
The reheating process is instantaneous for $w_{re}=\frac13$ and the reheating temperature
is at grand unification scale for this case. Hence parameters of reheating cannot be used for
constraining models of inflation.  
Now we  use  \eqrf{trenk} to obtain the final expression for $T_{re}$
\begin{equation}
T_{re}= \left[\left(\frac{43}{11g_{re}}\right)^{\frac{1}{3}} \frac{a_0 T_0}{c_S k} H_k \exp^{-N_k}\left[\frac{3^2.5 V_{end}}{\pi^2 g_{re}}\right]^{-\frac{1}{3(1+w_{re})}}\right]^{\frac{3(1+w_{re})}{3w_{re}-1}}.\label{tre1}
\end{equation}
The expressions for number of e-folds during reheating $N_{re}$, (\ref{nre1}), and reheating
temperature $T_{re}$, (\ref{tre1}), are the main results of this section. It is evident that
these two quantities depend on inflationary parameters $H_k$, $N_k$ and $V_{end}$, which can 
be expressed in terms of amplitude of scalar perturbations $A_s$ and spectral index $n_s$.
Hence bounds on reheating temperature and demanding $w_{re}$ to lie between $-\frac13$  and $1$
provide bounds on $n_s$. In subsequent sections we use these reheating parameters 
$N_{re}$ and $T_{re}$ to constrain noncanonical inflation with DBI kinetic term
and power-law kinetic term.

\section{ K-inflation with DBI kinetic term} \label{dbikinetic}
In this section  we consider K-inflation with DBI kinetic term, and monomial potentials and 
natural inflation potential.
 The Lagrangian for the scalar field in this case is given as
\begin{align}
{\cal L}  = - V(\phi)\sqrt{1-\eta^{2}g^{\mu\nu}\partial_{\mu}\phi \partial_{\nu}\phi}\bigg\}\label{action}.
\end{align}\\
Here $\eta$ has the dimension of $ [length]^2 $ and the field $\phi$ has the dimension of mass.
Using this Lagrangian we can obtain the energy density, (\ref{rhogen}), and 
pressure, (\ref{pgen}), for the background part 
of the scalar field in a homogeneous and isotropic universe as
\begin{align}
\rho=\frac{V(\phi)}{\sqrt{(1-\eta^2{\dot{\phi}}^2)}},\\
P=-V(\phi)(1-\eta^2{\dot{\phi}}^{2})^{\frac{1}{2}}. \label{rhonp}
\end{align}
Using \eqrf{rhonp} we can write the Friedmann equations for Hubble parameter  and its first
derivative as
\begin{align}
H^2 =\frac{1}{{3M_{P}}^2}\frac{V(\phi)}{(1-\eta^2{\dot{\phi}}^2)^{\frac{1}{2}}}\label{hsquare}, \\
\dot{H}=-\frac{V(\phi)\eta^2{\dot{\phi}}^2}{2M_{P}(1-\eta^2{\dot{\phi}}^2)^{\frac{1}{2}}}\label{hdot}.
\end{align}
The equation of motion for the background part of the scalar field can be
obtained from energy-momentum tensor (\ref{tmunu})   as
\begin{align}
\frac{\ddot{\phi}}{(1-\eta^2{\dot{\phi}}^2)}+3H\dot{\phi}+
\frac{V^\prime(\phi)}{\eta^2V(\phi)}=0\label{feqk}.
\end{align}
Here "$\prime$" refers to the derivative with respect to $\phi$. 
The Hubble flow parameters $\epsilon_1$ and $\epsilon_2$, 
for K-inflation with DBI kinetic term, can be obtained by substituting the expressions for
energy density and pressure (\ref{rhonp}) in \eqrf{ep1gen} and \eqrf{ep2gen} as
\bea
\epsilon_1 &=& \frac{3}{2}\eta^2\dot\phi^2,\label{eps1dbi}\\
\epsilon_2 &=& 2\frac{\ddot \phi}{H\dot\phi}.\label{etadbi}
\eea
Under slow-roll approximation $\ddot \phi$ in \eqrf{feqk} should be smaller than 
the friction term $3H\dot{\phi}$, and $\eta^2\dot\phi^2$ can be neglected in \eqrf{hsquare}.
Hence we obtain 
\be
\dot\phi = -\frac{V^\prime(\phi)}{3\eta^2HV(\phi)},\;\;H^2\sim\frac{V}{3M_P^2},\label{phidotfsq}
\ee
during inflation.
Using these approximations slow-roll parameters $\epsilon_1$ and $\epsilon_2$ can be
written in terms of the inflaton potential as
\bea
\epsilon_1 &=& \frac{M_P^2}{2}\left(\frac{V^{\prime2}}{\eta^2V^3}\right),\label{eps1dbiv}\\
\epsilon_2 &=& \frac{M_P^2}{\eta^2}
\left(-2\frac{V^{\prime\prime}}{V^2}+3\frac{V^{\prime 2}}{V^3}\right).\label{eps2dbiv} 
\eea
The amplitude of scalar perturbations $A_S$,  spectral index $n_s$ and tensor-to-scalar
ratio can now be obtained in terms of the parameters of inflaton potential using these
equations. Another parameter depending on inflaton potential is the number of e-foldings  
from the time when the Fourier mode $k$ leaves the Hubble radius during inflation to the end of 
inflation, which can be obtained using \eqrf{phidotfsq} as
\be
N_k=\int{H}dt=-\frac{\eta^2}{M_{P}^2}\int_{\phi_k}^{\phi_{end}}\frac{V^2}{V^{\prime}}d\phi
\label{nkdbi}
\ee

We now  impose reheating constraints on k-inflation having DBI kinetic term
with monomial  potential and PNGB potential.
\subsection{Monomial  potential}
We consider the following potential
\begin{align} 
V(\phi)= \frac{1}{2} m^{4-n}\phi^{n}.\label{polpot}
\end{align}
We choose $n =\frac{2}{3}\,,2\,$ and  $4$ for our analysis. This potential, for 
canonical single field inflation, in the context of reheating is studied in 
\cite{Martin:2013tda,Martin:2010kz,Dai:2014jja,Cook:2015vqa}.
Using \eqrftw{eps1dbiv}{eps2dbiv} for  potential  (\ref{polpot}), 
 the slow roll parameters can be obtained as
\bea
\epsilon_{1}&=&\frac{M_{P}^2 n^2}{\eta^2m^{4-n}\phi^{n+2}},\label{eps1pol}\\
\epsilon_2&=&\frac{2M_{P}^2 n (n+2)}{\eta^2m^{4-n}\phi^{n+2}}.\label{eps2pol}
\eea
At the end of inflation $\epsilon_1=1$ and hence the value of the scalar field at this time
can be obtained using \eqrf{eps1pol} as 
\begin{align}
\phi_{end}=\left(\frac{{M_{P}}^2 n^2}{\eta^2m^{4-n}}\right)^{\frac{1}{n+2}}\label{phiepol}.
\end{align}
The number of e-foldings  $N_k$ for monomial potential can be obtained using \eqrf{nkdbi}  
as 
\be
N_k=-\frac{\eta^2 m^{4-n}}{2M_{P}^2 n(n+2)}(\phi_{end}^{n+2}-\phi_k^{n+2}).\label{nkvpol}
\ee
Here $\phi_k$ is the value of inflaton field at the time when  mode $k$ leaves the horizon
during inflation. The spectral index $n_s$ can be obtained by substituting 
values of $\epsilon_1 $ and $\epsilon_2$ from \eqrf{eps1pol}, \eqrf{eps2pol} in \eqrf{nsk}
at $\phi = \phi_k $ as
\begin{align}
n_s = 1-\frac{4M_{P}^2 n(n+1)}{\eta^2m^{4-n}\phi_k^{n+2}}.\label{nspkpol}
\end{align}
Using this equation we get 
\begin{equation}
\phi_k = \left(\frac{4M_{P}^2 n(n+1)}{(1-n_s)\eta^2 m^{4-n}}\right)^{\frac{1}{n+2}},\label{phikpol}.
\end{equation}
and the slow-roll parameter $\epsilon_1$, (\ref{eps1pol}) at
 $\phi = \phi_k$ is given as 
\begin{equation}
\epsilon_1=\frac{n^2(1-n_s)}{4n(n+1)}.\label{epsnspol}
\end{equation}
Putting the values of $\phi_{end}$ and $\phi_k$ from \eqrf{phiepol} and \eqrf{phikpol} in \eqrf{nkvpol}, the number of e-foldings $N_k$ can be expressed in terms of spectral index $n_s$ as
\begin{equation}
N_k = \frac{n^2(3+n_s)+4n}{2n(n+2)(1-n_s)}.\label{polnk}
\end{equation}
 
The inflation potential at the end of inflation will be 
\begin{equation}
V_{end}=\frac{1}{2}m^{4-n}\phi_{end}^{n},\label{ve1}
\end{equation}
which can be expressed in terms of $H_k$ using \eqrf{phidotfsq} as
\begin{equation}
V_{end}=3M_{P}^2H_k^2\frac{\phi_{end}^n}{\phi_k^n}.\label{polvephi}
\end{equation}
Putting the values of $\phi_{end}$ and $\phi_k$ from \eqref{phiepol} and \eqref{phikpol} we
obtain
\begin{equation}
V_{end}=3M_{P}^2 H_{k}^2\bigg\{\frac{n^2(1-n_s)}{4n(n+1)}\bigg\}^{\frac{n}{n+2}}.\label{polvens}
\end{equation}
The  speed of sound $c_S$ for monomial potential with DBI kinetic term can be found using 
\eqrf{cs2} as 
\be
c_S = \sqrt{1-\frac{n^2\left(1-n_s\right)}{6n\left(n+1\right)}}\label{csnspol}.
\ee
The Hubble constant $H_k$ at the time when the mode $k$ leaves the horizon during inflation 
can be expressed in terms of scalar amplitude $A_S$ using \eqrf{scalar} as
\begin{equation}
H_k= \pi M_{P}\sqrt{8A_S \epsilon_1 c_S},\label{polhk}
\end{equation}
which can be written in terms of spectral index $n_s$ and $A_S$ using 
\eqrf{epsnspol} and \eqrf{csnspol} as
\begin{equation}
H_k=\pi M_{P}\sqrt{8A_S\sqrt{\bigg\{1-\frac{n^2(1-n_s)}{6n(n+1)}\bigg\}}\frac{n^2(1-n_s)}{4n(n+1)}}.\label{polhkns}
\end{equation}

\begin{figure}
\centering
\subfigure{ \includegraphics[width=7cm, height=6cm]{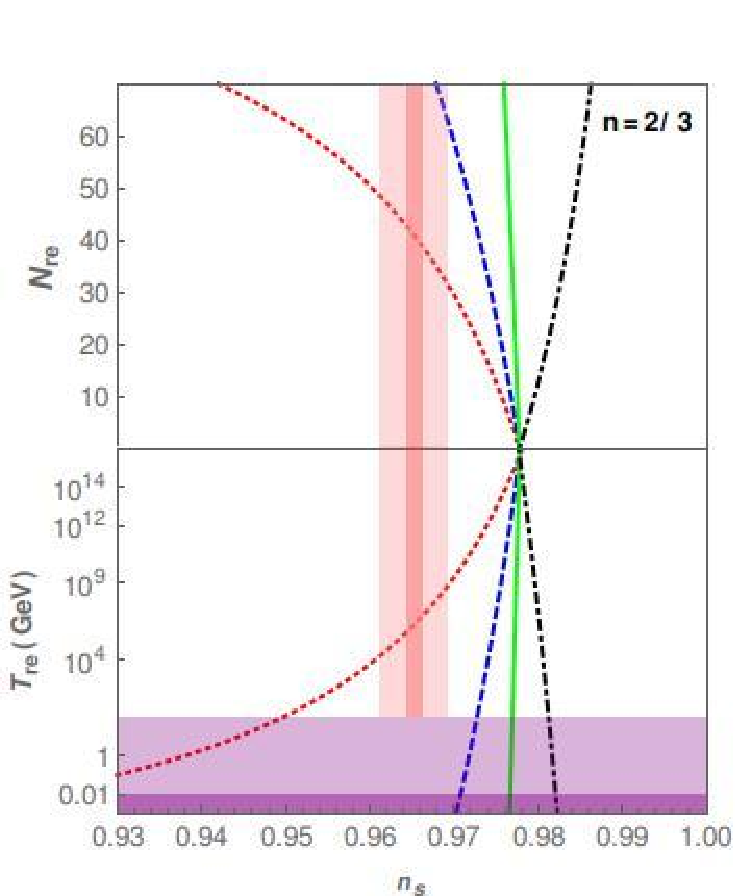}
        }
\subfigure{

          \includegraphics[width=7cm, height=6cm]{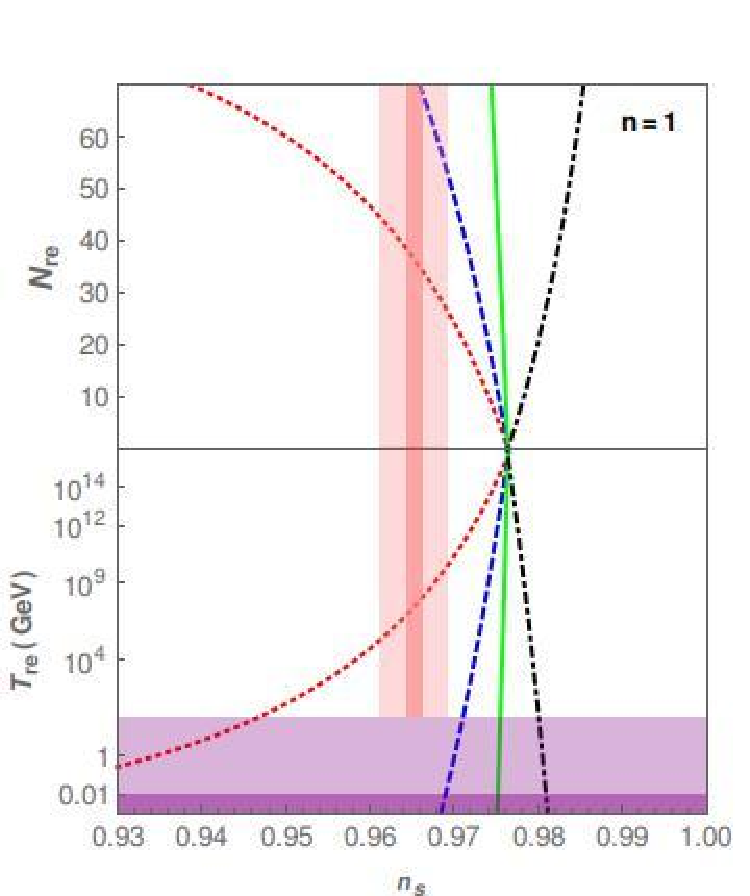}
        }
\subfigure{

            \includegraphics[width=7cm, height=6cm]{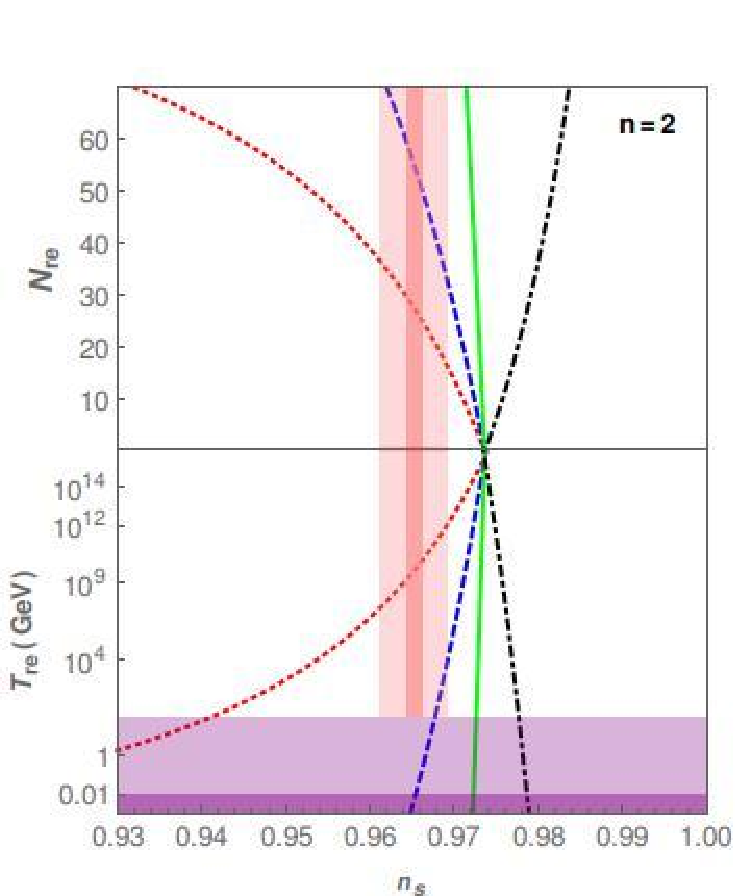}
        }
 \subfigure{

            \includegraphics[width=7cm, height=6cm]{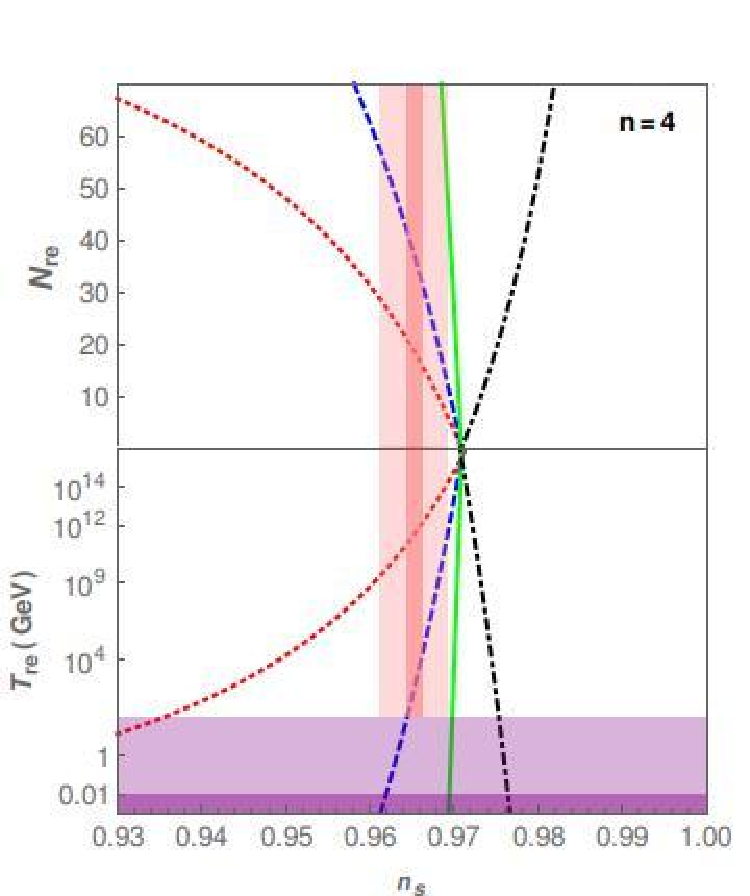}
        }

\caption{ $N_{re}$ and $ T_{re}$ as function of $n_s$ for four different values of $n$ of 
monomial potential. The vertical  pink region shows Planck-2018 bounds on $n_s$  
and dark pink region represents a precision of $10^{-3}$ from future observations 
\cite{Amendola:2016saw}.
The horizontal  purple region corresponds to $T_{re}$ of $10$ MeV from BBN and light
purple region corresponds to  100GeV of electroweak scale. 
Red dotted line corresponds to $w_{re} = -\frac{1}{3}$, 
blue dashed lines corresponds to $w_{re} = 0$, green solid line corresponds to  
$w_{re} = 0.25$ and  black dot-dashed line is for $w_{re}=1$.}
\label{fig:nretrepldbi}
\end{figure}

Using the expressions for  $N_k$ (\ref{polnk}), $V_{end}$ (\ref{polvens}) and 
$H_k$ (\ref{polhkns}), we can evaluate reheating temperature $T_{re}$ (\ref{tre1}) and 
e-folds during reheating $N_{re}$ in terms of spectral index for various equation of state.
Fig.~\ref{fig:nretrepldbi} depicts the variation of reheating temperature $T_{re}$ and 
$N_{re}$ with respect to $n_s$ for  $n=2/3$, $n=1$, $n=2$ and $n=4$.
We choose four values of effective equation of states during reheating 
$w_{re}=-1/3,\, 0,\, 0.25$ and $1$. The Planck-2018 bounds on $n_s=0.9853\pm0.0041$ are also 
shown in the figure. We have used Planck-2018 value  $A_S=2.20\times 10^{-9}$ for scalar
amplitude for our analysis. The point, where the curves of all $w_{re}$ meets, corresponds to 
instant reheating, $N_{re}\rightarrow 0$. The curve for $w_{re}$ would pass through this point 
and be vertical. 

By demanding that the reheating temperature should be above $100$ GeV for weak scale
dark matter production, we obtain bounds on 
spectral index by solving \eqrftw{tre1}{polnk}  and assuming
$-\frac13\le w_{re} \le 1$ for various choices of $n$.
These bounds on $n_s$ provides bounds on number of e-foldings $N_k$ from \eqrf{polnk}. 
The tensor-to-scalar ratio $r$ can be expressed in terms of $n_s$ using \eqrftw{rttsr}{epsnspol}
as
\begin{equation}
r=\frac{4n^2(1-n_s)}{n(n+1)}\bigg{[}1-\frac{n^2(1-n_s)}{6n(n+1)}\bigg{]}^{\frac{1}{2}}\label{polrns}.
\end{equation}
Using this expression the bounds on $n_s$, obtained using reheating temperature and effective
equation of state during reheating, can be transferred to the bounds on tensor-to-scalar ratio
$r$. 

\begin{table}
\begin{tabular}{ |c|c|c|c|c| } 
\hline
$ n $ &  Equation of state  & $ n_s $ & $ N_k $ & r  \\
\hline
\multirow{3}{4em}{$n = 2/3$}
& ${-1/3} \leq{w_{re}}\leq{0} $ & ${0.9497}\leq{n_s}\leq{0.9728} $ & ${24.72}\leq{N_k}\leq{45.79}$ & $ {0.0804}\geq{r}\geq{0.0435} $  \\ 
& ${0}\leq{w_{re}}\leq{0.25}  $ &  $ {0.9728}\leq{n_s}\leq{0.9769}$ & ${45.79}\leq{N_k}\leq{54.16}$ & ${0.0435}\geq{r}\geq{0.0368}  $  \\
& ${0.25}\leq{w_{re}}\leq{1}  $ & $ {0.9769}\leq{n_s}\leq{0.9813}$ & ${54.16}\leq{N_k}\leq{66.68}$ & $ {0.0368}\geq{r}\geq{0.0300} $  \\ 
\hline
\multirow{3}{4em}{$n=1$}
& $-1/3\leq{w_{re}}\leq{0}  $ & ${0.9468}\leq{n_s}\leq{0.9711}$ & $24.89\leq{N_k}\leq{45.97}$ & $ {0.1062}\geq{r}\geq{0.0577} $  \\ 
& ${0}\leq{w_{re}}\leq{0.25}$ & ${0.9711}\leq{n_s}\leq{0.9755}$ & $45.97\leq{N_k}\leq{54.34}$  & ${0.0577}\geq{r}\geq{0.0489}  $ \\
& ${0.25}\leq{w_{re}}\leq{1}$ & $ {0.9755}\leq{n_s}\leq{0.9801}$ & $54.34\leq{N_k}\leq{66.84}$ & $ {0.0489}\geq{r}\geq{0.0398} $  \\ 
\hline
\multirow{3}{4em}{$n=2$} 
&  $-1/3\leq{w_{re}}\leq{0} $  & ${0.9411}\leq{n_s}\leq{0.9678}$  & $25.21\leq{N_k}\leq{46.28}$ & $ {0.1566}\geq{r}\geq{0.0858} $ \\
&  ${0}\leq{w_{re}}\leq{0.25}$ &  ${0.9678}\leq{n_s}\leq{0.9727}$ & $46.28\leq{N_k}\leq{54.63}$ & ${0.0858}\geq{r}\geq{0.0728}$   \\ 
& ${0.25}\leq{w_{re}}\leq{1} $ & $ {0.9727}\leq{n_s}\leq{0.9777}$ & $54.63\leq{N_k}\leq{67.11}$ & $ {0.0728}\geq{r}\geq{0.0593}$ \\ 
\hline
\multirow{3}{4em}{$n=4$} 
& $-1/3\leq{w_{re}}\leq{0} $   & ${0.9355}\leq{n_s}\leq{0.9645}$ & $ 25.50\leq{N_k}\leq{46.56}$ & ${0.2055}\geq{r}\geq{0.1135} $ \\ 
& ${0}\leq{w_{re}}\leq{0.25}$  & ${0.9645}\leq{n_s}\leq{0.9698}$ & $ 46.56\leq{N_k}\leq{54.89}$ & ${0.1135}\geq{r}\geq{0.0963}$  \\ 
& ${0.25}\leq{w_{re}}\leq{1}$ & $ {0.9698}\leq{n_s}\leq{0.9754}$ & $ 54.89\leq{N_k}\leq{67.33}$ & $ {0.0963}\geq{r}\geq{0.0786}$  \\  
\hline
\end{tabular}
\caption{The allowed values of spectral index $n_s$ and number of e-folds $N_k$ for various values of
 $n$ for monomial potential by demanding $T_{re}\geq{100GeV}$.}
\label{table:nsnkrpoldbi}
\end{table}
\begin{figure}[!h]
\centering

        \subfigure[$N_k$ vs $n_s$ plot for $n = 2/3, 1, 2, 4$]{
           
          \includegraphics[width=7cm, height=5.5cm]{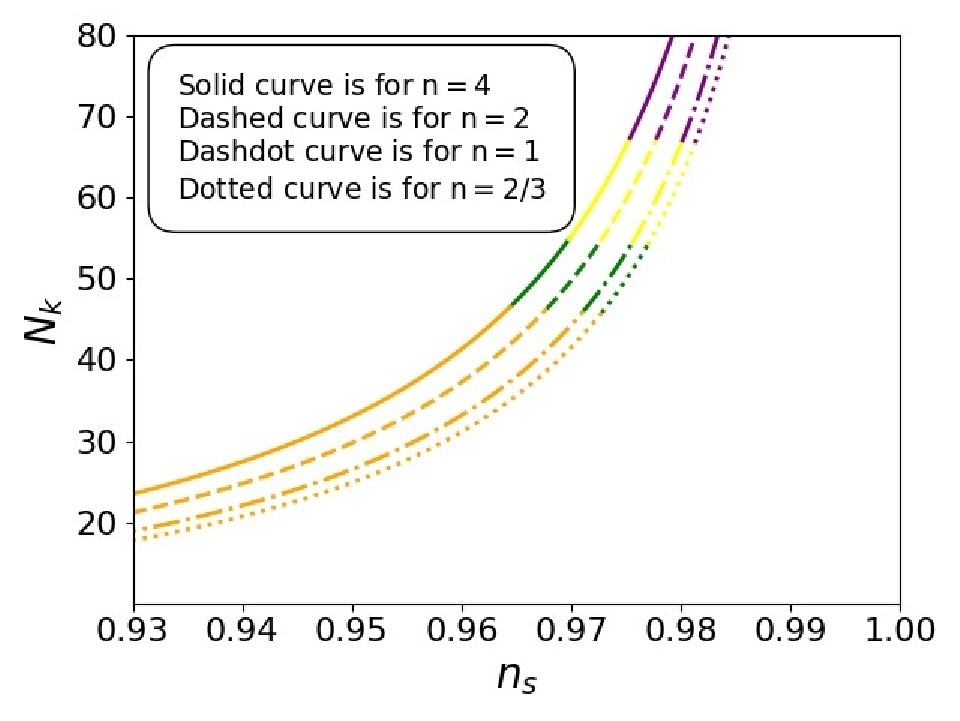}
        }
        \subfigure[$r$ vs $n_s$ plot for $n = 2/3, 1, 2, 4$]{
            
            \includegraphics[width=7cm, height=5.5cm]{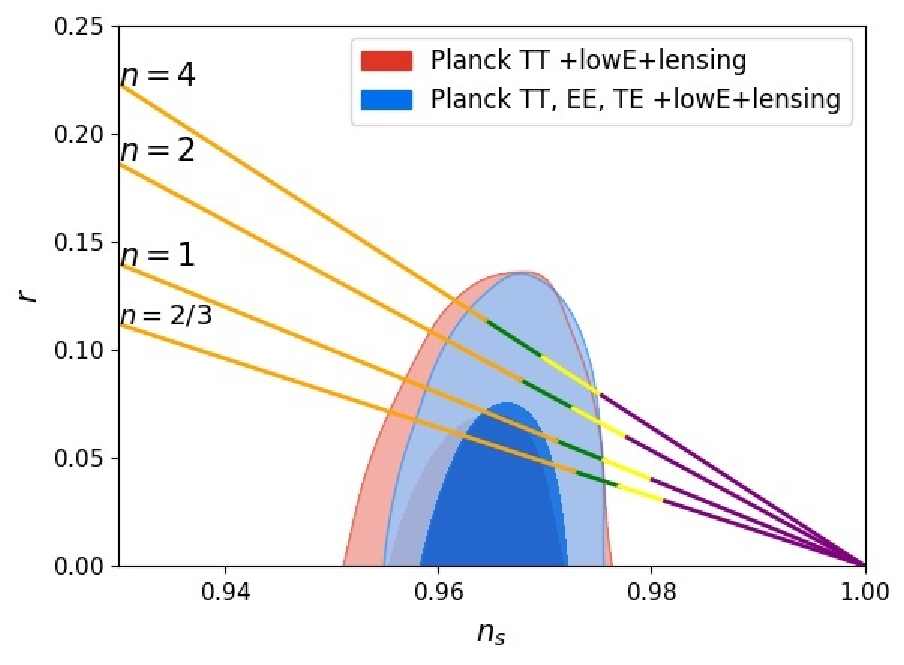}
        }
           
\caption{  $N_k$ vs $n_s$, and $r-n_s$ predictions along with joint $68\%$ and $95\%$ C.L. 
Planck-2018 constraints for monomial potentials with DBI kinetic term. 
Here in both panels the orange region corresponds to $w_{re}\le 0$, green region corresponds to 
$0\le w_{re}\le 0.25$, yellow region shows $0.25\le w_{re}\le 1$ and purple region corresponds 
to $w_{re}>1$.}
\label{fig:nkrnspoldbi}
\end{figure}
The bounds on $n_s$, $N_k$ and $r$, thus obtained, are listed in Table~\ref{table:nsnkrpoldbi}.
It can be seen from Table ~\ref{table:nsnkrpoldbi} and Fig.~\ref{fig:nretrepldbi}  
that, for $n=2/3$ and $1$, the bounds on $n_s$ lies outside the Planck-2018 bounds, if we demand that the effective
equation of state lie between the physically plausible range $0\le w_{re}\le 0.25$.
With this range of $w_{re}$ the tensor-to-scalar ratio $r$ for quadratic and quartic 
potential is slightly greater than joint BICEP2/Keck Array and Planck bounds $r<0.06$ 
\cite{Ade:2018gkx}.

The plots between $N_k$ and $n_s$ are shown in the left panel of Fig.~\ref{fig:nkrnspoldbi} for various values of
$n$ and $w_{re}$. The tensor-to-scalar ratio $r$ as a function $n_s$ for the four choices
of monomial potentials, is shown in right panel of Fig.~\ref{fig:nkrnspoldbi} along with 
joint  68\% and 95\% C.L constraints from Planck-2018. It can be seen from Fig.~\ref{fig:nkrnspoldbi} that $r$ vs $n_s$ predictions
for the quadratic and quartic potential with DBI kinetic term lie within 95\% C.L. but lie
outside 68\% C.L. of Planck-2018 data for physically plausible range of $0\le w_{re}\le 0.25$.
However, potential with $n=\frac{2}{3}$ and $n=1$ lie well within 68\% of Planck-2018
observations, but, for this the equation of state during reheating should be less than $0$.

\subsection{Natural inflation potential}
\begin{figure}[h]
\centering
        \subfigure{
           
          \includegraphics[width=7cm, height=6cm]{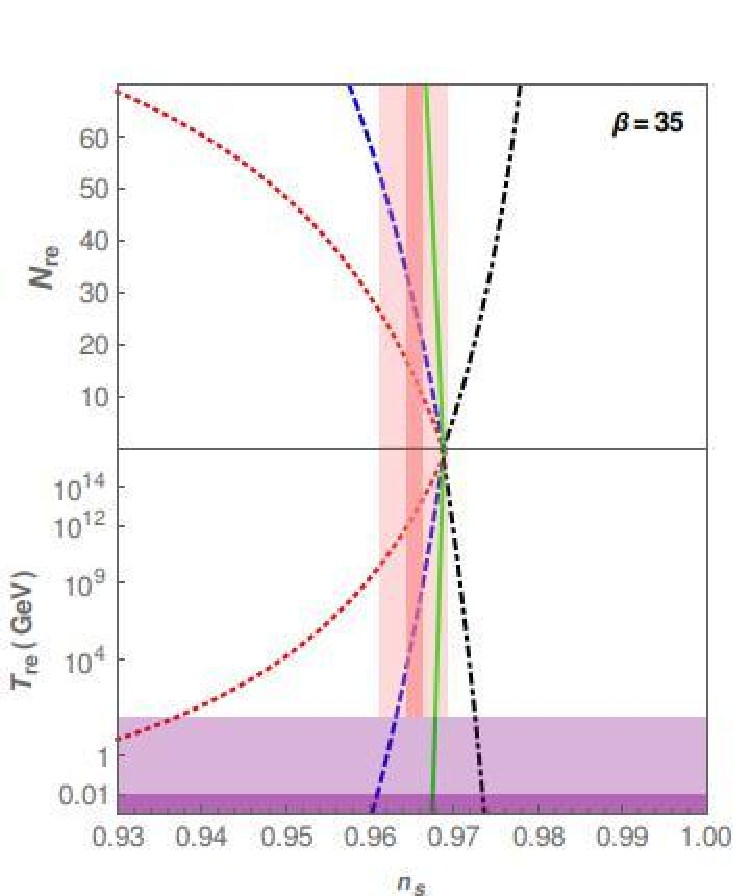}
        }
        \subfigure{
           
          \includegraphics[width=7cm, height=6cm]{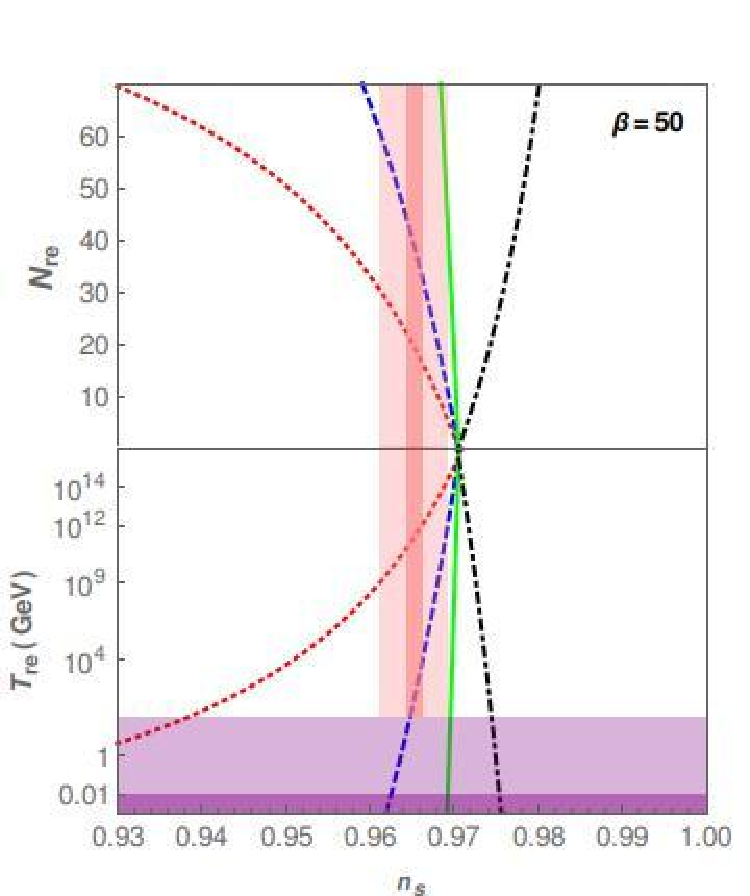}
        }
        \subfigure{
            
            \includegraphics[width=7cm, height=6cm]{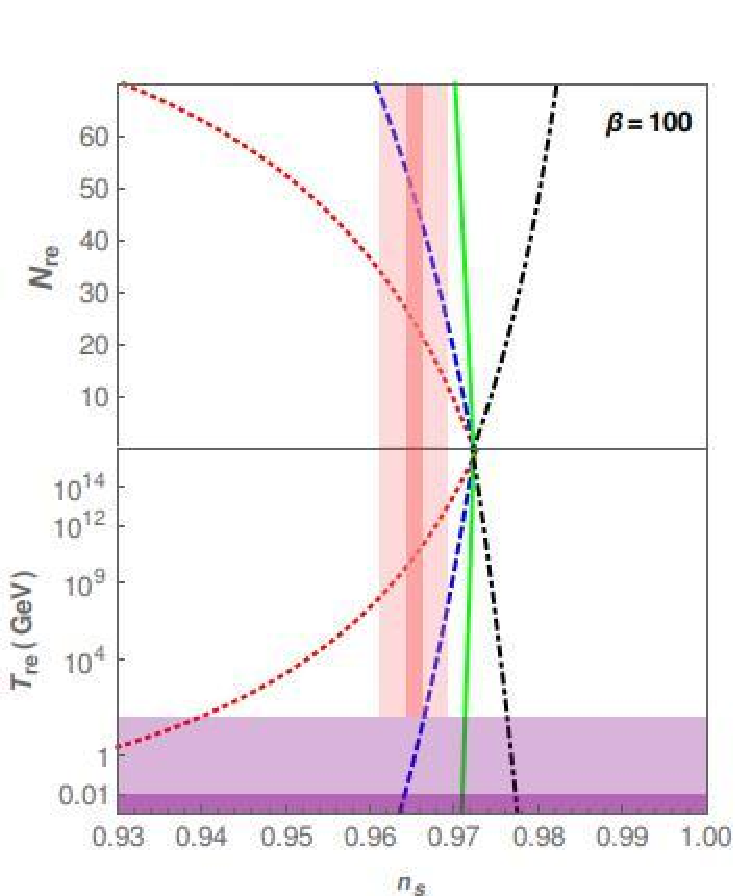}
        }
        \subfigure{
            
            \includegraphics[width=7cm, height=6cm]{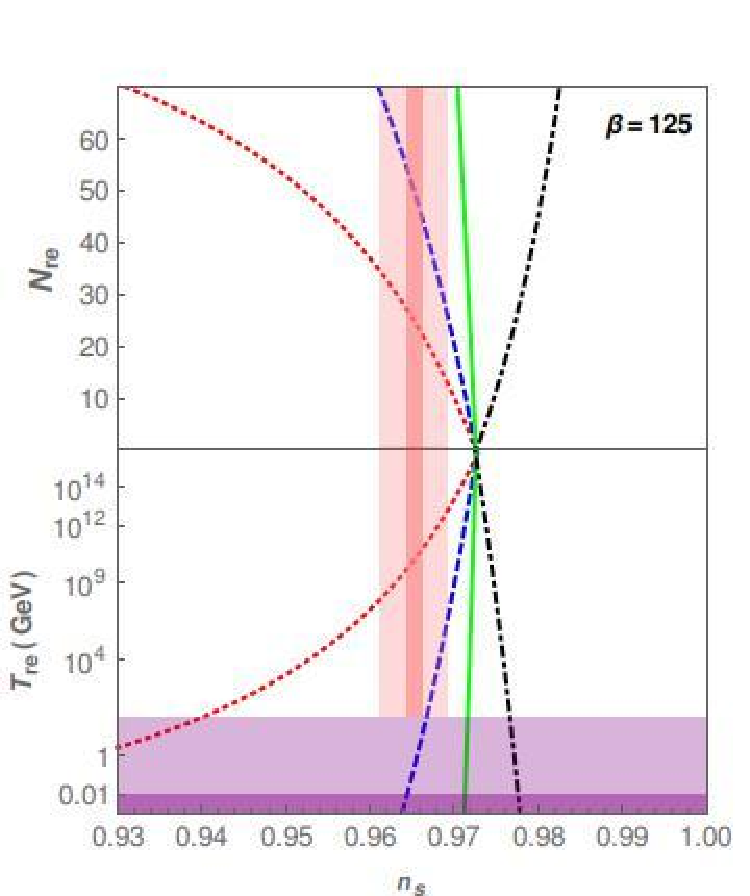}
        }
\caption{$N_{re}$ and $ T_{re}$ as function of $n_s$ for 
natural inflation potential. The vertical  pink region shows Planck-2018 bounds on $n_s$  
and dark pink region represents a precision of $10^{-3}$ from future observations 
\cite{Amendola:2016saw}.
The horizontal  purple region corresponds to $T_{re}$ of $10$ MeV from BBN and light
purple region corresponds to  100GeV of electroweak scale. 
Red dotted line corresponds to $w_{re} = -\frac{1}{3}$, 
blue dashed lines corresponds to $w_{re} = 0$, green solid line corresponds to  
$w_{re} = 0.25$ and  black dot-dashed line is for $w_{re}=1$.}
\label{fig:nretrenatdbi}
\end{figure}

The potential for Pseudo-Nambu-Goldstone-Boson, natural inflation  is given as 
\cite{Freese:1990rb}
\begin{equation}
V(\phi)=\Lambda^4\bigg{[}1+\cos\bigg{(}\frac{\phi}{f}\bigg{)}\bigg{]},\label{potn}
\end{equation}
where $f$ is the spontaneous symmetry breaking scale and $\Lambda$ is explicit symmetry 
breaking scale for pseudo-Nambu-Goldstone boson.
Reheating constraints on this potential with noncanonical kinetic term having DBI form are 
discussed in \cite{Rashidi:2017chf}. Here we revisit these constraints with Planck-2018 data.
  Defining $\beta\equiv\eta^2f^2\Lambda^4 M_{P}^{-2}$ the slow-roll 
parameters for potential given in \eqrf{potn} can be obtained using  \eqrftw{eps1dbiv}{eps2dbiv}
as
\bea
\epsilon_1&=&\frac{1}{2\beta}\frac{1-\cos\bigg{(}\frac{\phi}{f}\bigg{)}}{\bigg\{1+\cos\bigg{(}\frac{\phi}{f}\bigg{)}\bigg\}^2},\label{eps1beta}\\
\epsilon_2&=&\frac{1}{\beta} \frac{3-\cos\bigg{(}
\frac{\phi}{f}\bigg{)}}{\bigg\{1+\cos\bigg{(}\frac{\phi}{f}\bigg{)}\bigg\}^2}\label{epsn2}.
\eea
The value of inflaton field at the end of inflation can be obtained 
by setting $\epsilon_1=1$ as
\begin{equation}
\cos\bigg{(}\frac{\phi_{end}}{f}\bigg{)}=\frac{-(4\beta+1)+\sqrt{(1+16\beta)}}{4\beta}.\label{cosphi}
\end{equation}
The  spectral index  $n_s$ can be obtained by substituting values of 
$\epsilon_1$ (\ref{eps1beta}) and $\epsilon_2$ (\ref{epsn2}) in \eqrf{nsk} as
\begin{align}
n_s&=1-\frac{1}{\beta}\frac{1-\cos\bigg{(}\frac{\phi}{f}\bigg{)}}{\bigg\{1+\cos\bigg{(}\frac{\phi}{f}\bigg{)}\bigg\}^2}-\frac{1}{\beta} \frac{3-\cos\bigg{(}\frac{\phi}{f}\bigg{)}}{\bigg\{1+\cos\bigg{(}\frac{\phi}{f}\bigg{)}\bigg\}^2}\label{nscos}\\
  &=1-\frac{2\bigg{[}2-\cos\bigg{(}\frac{\phi}{f}\bigg{)}\bigg{]}}{\beta\bigg{[}1+\cos\bigg{(}\frac{\phi}{f}\bigg{)}\bigg{]}^2}.\label{nscosnp}
\end{align}
Number of e-foldings for potential (\ref{potn}) can be expressed using \eqrf{nkdbi} as
For natural inflation potential \eqrf{potn} , $N_k$ can be written as:-
\bea
N_k&=&\frac{\beta}{f}\int_{\phi_k}^{\phi_{end}}\frac{\bigg{[}1+\cos\big{(}\frac{\phi}{f}\big{)}\bigg{]}^2}{\sin\bigg{(}\frac{\phi}{f}\bigg{)}}\nonumber\\
&=&\beta\bigg{[}\cos\big{(}\frac{\phi_{end}}{f}\big{)}-\cos\big{(}\frac{\phi_k}{f}\big{)}\bigg{]}+2\beta\ln{\bigg{[}\frac{\cos\big{(}.\frac{\phi_{end}}{f}\big{)}-1}{\cos\big{(}\frac{\phi_k}{f}\big{)}-1}\bigg{]}},\label{nkphinp}
\eea
where again $\phi_{end}$ and $\phi_k$ are the values of inflaton field at the end of inflation
and at the time the mode $k$ leaves inflationary horizon during inflation respectively.
Defining  $\cos\big{(}\frac{\phi_{end}}{f}\big{)}=x$ and  $\cos\big{(}\frac{\phi_k}{f}\big{)}=y$,
 \eqrf{nkphinp} for number of e-folds $N_k$ can be written  as 
\be
N_k=\beta x -\beta y+2 \beta \ln{(x-1)}-2\beta\ln{(y-1)}. \label{nkxy}
\ee
 The spectral index $n_s$, (\ref{nscosnp}), at $\phi=\phi_k$   will have the form in terms of $y$
as
\begin{align}
n_s= 1-\frac{2}{\beta}\frac{(2-y)}{(1+y)^2}.\label{nsnp}
\end{align}
To express $N_k$ in terms of $n_s$, \eqrf{nsnp} can be solved for $y$ as
\begin{equation}
y=1+\frac{1+2\beta-2n_s\beta-\sqrt{1+6\beta-6n_s\beta}}{n_s\beta-\beta},\label{yns}
\end{equation}
and $x$ is given by \eqrf{cosphi}.
Inflaton potential at the end of inflation can be given as
\begin{equation}
V_{end}=\Lambda^4\bigg{[}1+\cos\big{(}\frac{\phi_{end}}{f}\big{)}\bigg{]},\label{vephinp}
\end{equation}
which can be written using \eqrf{phidotfsq} as
\bea
V_{end}&=&3M_{P}^2 H_k^2\frac{\bigg{[}1+\cos\big{(}\frac{\phi_{end}}{f}\big{)}\bigg{]}}{\bigg{[}1+\cos\big{(}\frac{\phi_{k}}{f}\big{)}\bigg{]}}\nonumber\\
&=&3M_{P}^2 H_k^2\frac{(1+x)}{(1+y)}.\label{vexy}
\eea
From \eqrf{cs2}, the speed of sound $ c_S$ at $\phi=\phi_k$ can be written as
\begin{align}\nonumber
c_S&=\sqrt{1-\frac{1}{3\beta}\frac{(1-\cos\big{(}\frac{\phi_k}{f}\big{)})}{(1+\cos\big{(}\frac{\phi_k}{f}\big{)})^2}},\\
   &=\sqrt{1-\frac{1}{3\beta}\frac{(1-y)}{(1+y)^2}}.\label{csxy}
\end{align}
The value of the Hubble constant at the time when Fourier mode $k$ leaves the inflationary 
horizon during inflation can again be expressed in terms of amplitude of 
scalar perturbations $A_S$ by putting the values of $\epsilon_1$ 
(\ref{eps1beta}), and $c_S$, (\ref{csxy}), in \eqrf{scalar} as
\begin{equation}
H_k=\pi M_{P}\sqrt{8A_S \frac{1}{2\beta}\frac{1-y}{(1+y)^2}\sqrt{1-\frac{1}{3\beta}\frac{(1-y)}{(1+y)^2}}}.\label{hknp}
\end{equation}
We can express $ N_k $, $V_{end}$ and $H_k$ in terms of spectral index by substituting the value
of $y$ from \eqrf{yns} and $x$ from \eqrf{cosphi} in  \eqrf{nkxy}, \eqrf{vexy} and  \eqrf{hknp},
and then using these expressions the reheating temperature $T_{re}$ and  number of e-folds
during reheating $N_{re}$ can be obtained in terms of spectral index  from
\eqrftw{nre1}{tre1}. 
We have chosen  $\beta=35, 50, 100$ and $125$ for our analysis. Increasing $\beta$ beyond
$125$ does not affect the results.
The variation of $N_{re}$ and $T_{re}$ with respect to $n_s$, along with 
Planck-2018 bounds on $n_s=0.9853\pm 0.0041$, is represented in 
Fig.~\ref{fig:nretrenatdbi} for various values of effective equation state during 
reheating. Again the curves for various values of $w_{re}$ meet at the point corresponding to
instant reheating, $N_{re}\rightarrow 0$. The curve for $w_{re}=1/3$ would pass through this
point and be vertical.

By imposing the bounds  on $T_{re}$, i.e., $T_{re}>100$ GeV for weak scale dark matter production,
we obtain bounds on $n_s$ for various equation of states $w_{re}$ solving \eqrf{tre1}.
Again the tensor-to-scalar ration for natural inflation with DBI kinetic term can be
obtained from \eqrftr{tensor}{eps1beta}{csxy} as
\be
r = \frac{8}{2\beta}\frac{\left(1-y\right)}{\left(1+y\right)^2}\sqrt{1-\frac{1}{3\beta}\frac{(1-y)}{(1+y)^2}}
\ee

The bounds on $n_s$ obtained from $T_{re}$ and $w_{re}$ can give bounds on $N_k$ and $r$.
These bounds for various choices of $\beta$ are given in Table~\ref{table:nsnkrnatdbi}
\begin{table}[h]
\begin{tabular}{ |c|c|c|c|c|} 
\hline
$\alpha $ &  Equation of state  & $ n_s $& $ N_k $ & r  \\
\hline
\multirow{3}{4em}{$\beta = 35 $}
& $-1/3\leq{w_{re}}\leq{0} $& ${0.9369}\leq{n_s}\leq{0.9631}$ &$25.08\leq{N_k}\leq{46.06}$& $ {0.1072}\geq{r}\geq{0.0480} $  \\ 
& ${0}\leq{w_{re}}\leq{0.25}$ &  ${0.9631}\leq{n_s}\leq{0.9678}$ &$46.06\leq{N_k}\leq{54.37}$& ${0.0480}\geq{r}\geq{0.0377}$  \\
&${0.25}\leq{w_{re}}\leq{1} $& $ {0.9678}\leq{n_s}\leq{0.9725}$ &$54.37\leq{N_k}\leq{66.78}$&$ {0.0377}\geq{r}\geq{0.0274}$  \\ 
\hline
\multirow{3}{4em}{$\beta=50$}
&$-1/3\leq{w_{re}}\leq{0}  $ & ${0.9384}\leq{n_s}\leq{0.9648}$ &$25.10\leq{N_k}\leq{46.10}$ &$ {0.1159}\geq{r}\geq{0.0547} $\\ 
&${0}\leq{w_{re}}\leq{0.25}$ & ${0.9648}\leq{n_s}\leq{0.9696}$ &$46.10\leq{N_k}\leq{54.42}$ & ${0.0547}\geq{r}\geq{0.0438}$ \\
&${0.25}\leq{w_{re}}\leq{1}$ & ${0.9696}\leq{n_s}\leq{0.9745}$ & $54.42\leq{N_k}\leq{66.86}$ &$ {0.0438}\geq{r}\geq{0.0329}$ \\ 
\hline
\multirow{3}{4em}{$\beta=100$} 
&  $-1/3\leq{w_{re}}\leq{0} $& ${0.9398}\leq{n_s}\leq{0.9665}$  &$25.13\leq{N_k}\leq{46.16}$ &$ {0.1287}\geq{r}\geq{0.0645} $\\
& ${0}\leq{w_{re}}\leq{0.25}$ &  ${0.9665}\leq{n_s}\leq{0.9713}$ &$46.16\leq{N_k}\leq{54.49}$& ${0.0645}\geq{r}\geq{0.0529}$  \\ 
& ${0.25}\leq{w_{re}}\leq{1} $& $ {0.9713}\leq{n_s}\leq{0.9764}$ &$54.49\leq{N_k}\leq{66.95}$ &$ {0.0529}\geq{r}\geq{0.0412}$ \\ 
\hline
\multirow{3}{4em}{$\beta=125$} 
& $-1/3\leq{w_{re}}\leq{0} $& ${0.9401}\leq{n_s}\leq{0.9668}$ & $25.14\leq{N_k}\leq{46.18}$  &$ {0.1318}\geq{r}\geq{0.0669} $ \\ 
& ${0}\leq{w_{re}}\leq{0.25}$ &  ${0.9668}\leq{n_s}\leq{0.9717}$ &$46.18\leq{N_k}\leq{54.51}$ & ${0.0669}\geq{r}\geq{0.0552}$  \\ 
& ${0.25}\leq{w_{re}}\leq{1} $& $ {0.9717}\leq{n_s}\leq{0.9767}$ &$54.51\leq{N_k}\leq{66.97}$&$ {0.0552}\geq{r}\geq{0.0432}$  \\  
\hline
\end{tabular}
\caption{The allowed values of spectral index $n_s$ and number of e-folds $N_k$ for various 
values of  $\beta$ for natural inflation potential, obtained by imposing $T_{re}\geq{100GeV}$.}
\label{table:nsnkrnatdbi}
\end{table}

\begin{figure}[!htp]
\centering
\subfigure[$N_k$ vs $n_s$ plot for $ \beta < 125$]{
 \includegraphics[width=6cm, height=5cm]{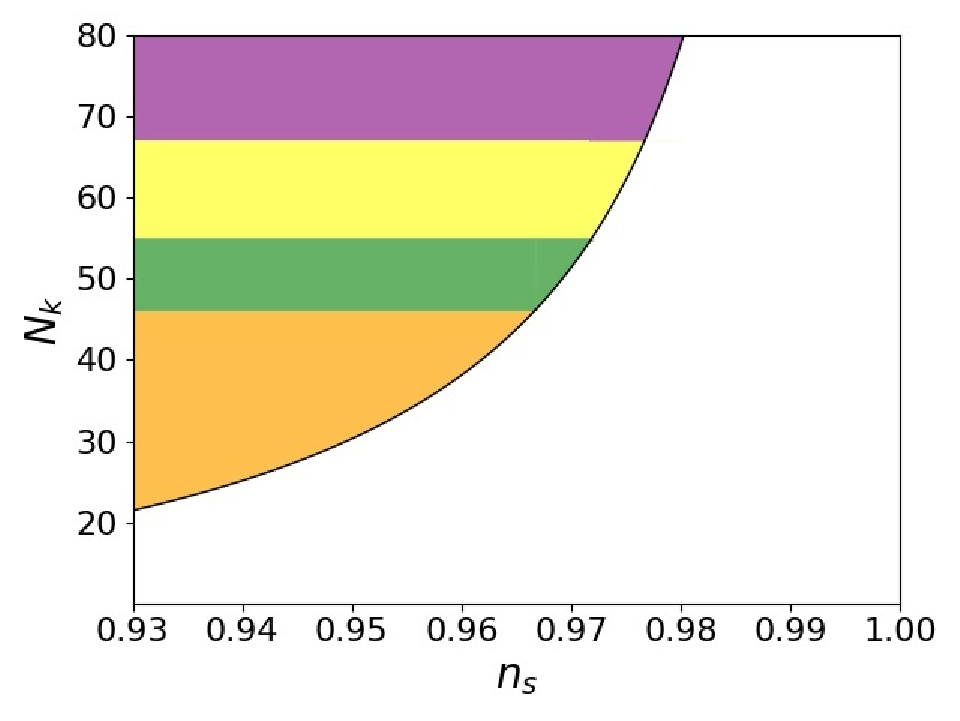}
        }
        \subfigure[r vs $n_s$ plot for $ \beta < 125$]{
           
          \includegraphics[width=6cm, height=5cm]{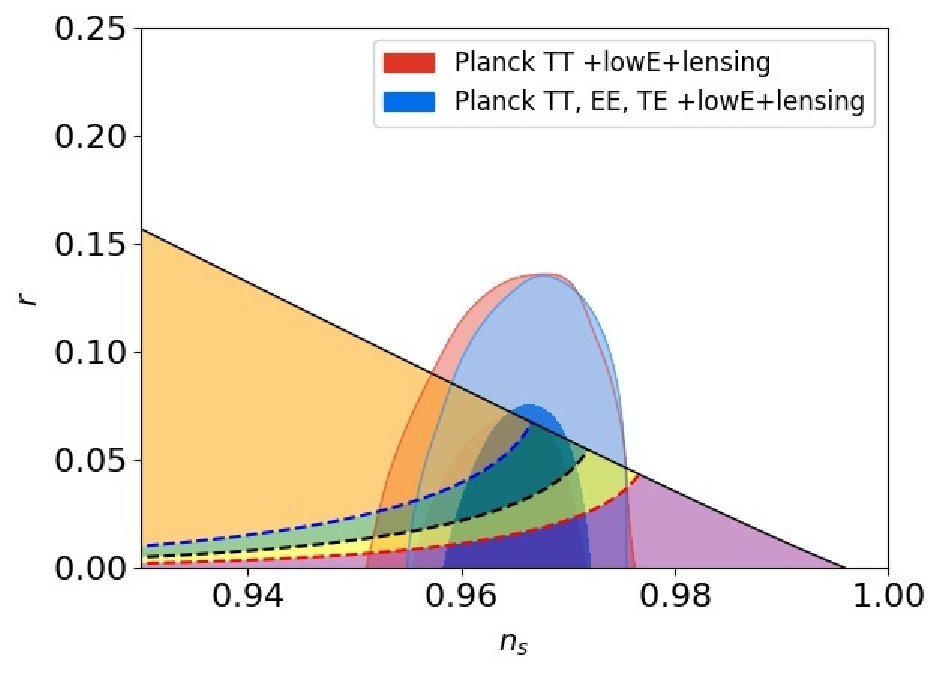}
        }
       
\caption{  $N_k$ vs $n_s$ and $r$ vs $n_s$ plots for natural inflation potential along with joint 68\%CL and95\%CL Planck-2018 constraints. In both the panels the orange region corresponds to 
$w_{re}< 0$, the green region corresponds to $0< w_{re}<0.25$, the yellow region corresponds to 
$0.25< w_{re}<1$ and the purple region corresponds to $w_{re}>1$. 
In the right panel of the figure  blue dashed  line corresponds to $N_k= 46$,  
black dashed  line corresponds to $N_k= 55$ and  red dashed line corresponds to $N_k= 67$. 
These values of $N_k$  corresponds to bounds on $n_s$ obtained by demanding $T_{re}>100 $ GeV 
for different values of $w_{re}$. The solid  black line in  both the panels of the figure  
corresponds to $\beta =125$ and the filled region corresponds to $\beta < 125$.}
\label{fig:nkrnsnatdbi}
\end{figure}  

It can be seen from  Table~\ref{table:nsnkrnatdbi} that  with physically plausible range  $0\le w_{re}\le 0.25$
the bounds on $n_s$ and $r$ are compatible with Planck-2018 observations  for $\beta < 125$.
We also show $N_k$ vs $n_s$  and $r$ vs $n_s$ plots for PNGB potential with DBI kinetic term
in Fig.~\ref{fig:nkrnsnatdbi}. It is evident from the figure that the
values of $n_s$ and $r$ predicted in this model lie within $1\sigma$ contour of Planck-2018
joint constraints for physically plausible range $0\le w_{re}\le 0.25$ shown by green region in
the figure. Our results for natural inflation with DBI kinetic term agree with
\cite{Rashidi:2017chf}.

\section{K-inflation with power-law kinetic term} \label{powerlawkinetic}
In this section we will analyze K-inflation with power-law kinetic term. The Lagrangian density
for this case is given as \cite{Mukhanov:2005bu,Unnikrishnan:2012zu}
\begin{equation}
L(X,\phi) = X\left(\frac{X}{M^4}\right)^{\alpha-1} - V(\phi),\label{lagnc}
\end{equation}
where M has dimension of mass and $\alpha$ is dimensionless. 
For $\alpha=1$ the Lagrangian reduces to usual canonical scalar field. 
Using \eqrftw{rhogen}{pgen} the energy density and pressure can be obtained as
\begin{equation}
\rho_{\phi}=(2\alpha-1) X\left(\frac{X}{M}\right)^{\alpha-1} + V(\phi).\label{rhonc}
\end{equation}
\begin{equation}
p_{\phi}=X\left(\frac{X}{M}\right)^{\alpha-1}-V(\phi),  X\equiv\frac{1}{2}\dot{\phi}^2.\label{pnc}
\end{equation}
Thus, the  Friedman equations for Hubble constant and its first derivative  become
\bea
H^2 &=& \frac{8\pi G}{3}\left[(2\alpha-1)X\left(\frac{X}{M^4}\right)^{\alpha-1} + V(\phi)\right],\label{h2nc}\\
\dot{H} &=& -4\pi G(\rho_{\phi}+p_{\phi}) = -\frac{1}{3M_{P}^2} X\left(\frac{X}{M^4}\right)^{\alpha-1}.\label{Hdot}
\eea

The evolution equation for inflaton $\phi$ can be obtained by energy-momentum tensor 
(\ref{tmunu}) as
 
\begin{equation}
\ddot{\phi}+\frac{3H\dot{\phi}}{2\alpha-1}+\left(\frac{V'(\phi)}{\alpha(2\alpha-1)}\right)\left(\frac{2M^4}{\dot{\phi^2}}\right)^{\alpha-1}=0.\label{eqmnc}
\end{equation}
Using the definition of slow-roll parameter $\epsilon=-\dot H/H^2$, along with \eqrf{Hdot}, 
the Hubble constant (\ref{h2nc}) can be written as
\begin{equation}
H^2\left[1-\left(\frac{2\alpha-1}{3\alpha}\right)\epsilon\right] = \frac{1}{3M_{P}^2}V(\phi),\label{frw}
\end{equation}
under slow-roll approximation $\epsilon << 1$ this reduces to
\begin{equation}
H^2 = \frac{V(\phi)}{3M_{P}^2}\label{Haprox}.
\end{equation}
For slow-roll  $\ddot{\phi}$ is much smaller than the friction term in \eqrf{eqmnc}, 
hence using \eqrf{Haprox}, we obtain
\begin{equation}
\dot{\phi}=  \left[\left(\frac{M_{P}}{\alpha \sqrt{3}}\right)\left(\frac{-V'(\phi)}{\sqrt{V}}\right)(2M^4)^{\alpha-1}\right]^{\frac{1}{2\alpha-1}}\label{phidotnc}.
\end{equation}
The two Hubble flow parameters $\epsilon_1$, (\ref{ep1gen}), and $\epsilon_2$, 
(\ref{ep2gen}), for this case can be obtained using 
Eqns~(\ref{rhonc}), (\ref{pnc}), (\ref{Haprox}) and  \eqrf{phidotnc}, as
\begin{equation}
\epsilon_1 = \left[\frac{1}{\alpha}\left(\frac{3M^4}{V}\right)^{\alpha-1}\left(\frac{-M_{P} V'}{\sqrt{2}V}\right)^{2\alpha}\right]^{\frac{1}{2\alpha-1}}\label{eps1nc},
\end{equation}
\begin{equation}
\epsilon_2 = \frac{-2\epsilon_1}{2\alpha-1}\left[2\alpha\left(\frac{V'' V}{V'^2}\right) - (3\alpha-1)\right]\label{eps2nc}.
\end{equation}
Now the number of e-foldings $N_k$ from the time when mode k leaves the horizon to the end of
inflation, in case of power-law kinetic term, can be obtained by using
\begin{equation}
N_k = -\int_{\phi_{end}}^{\phi_k}\left(\frac{H}{\dot\phi}\right) d\phi,\label{efoldch}
\end{equation}
and substituting the values of $H$ and $\dot \phi$  from \eqrftw{Haprox}{phidotnc} respectively in
this expression for various choices of potentials.
The speed of sound $c_S$, defined in \eqrf{cs2}, 
can be obtained using \eqrf{rhonc} and \eqrf{pnc} as
\begin{equation}
c_S^2 = \frac{1}{2\alpha-1}\label{cs2nc}.
\end{equation}
The speed of sound here  is only function of $\alpha$ and independent of choice of potential.

\subsection{Monomial potentials}
We consider following monomial potential with power-law kinetic term
\begin{equation}
V(\phi)= \frac{1}{2}m^{4-n} \phi^n , \quad where \quad  n > 0.\label{chaopot}
\end{equation}
The two Hubble-flow parameters for this potential can be obtained  using
\eqrftw{eps1nc}{eps2nc}
\bea
\epsilon_1 &=& \left[\frac{1}{\alpha}\left(\frac{6M^4}{m^{4-n}}\right)^{(\alpha-1)}\left(\frac{-n M_{P}}{\sqrt{2}}\right)^{2\alpha}\frac{1}{\phi^{2\alpha+n\alpha-n}}\right]^{\frac{1}{2\alpha-1}},
\label{eps1ch}\\
\epsilon_2 &=& \frac{2\epsilon_1 \gamma}{n}\label{eps2ch2}.
\eea
Here
\begin{equation}
\gamma \equiv \frac{2\alpha+n(\alpha-1)}{2\alpha-1}.\label{gamma}
\end{equation} 
The value of the inflaton field at  the end of inflation, $\phi_{end}$, can be obtained by 
setting   $ \epsilon_1 =1$ as
\begin{equation}
\phi_{end} = \left[\frac{1}{\alpha}\left(\frac{6M^4}{m^{4-n}}\right)^{(\alpha-1)} \left(\frac{-nM_{P}}{\sqrt{2}}\right)^{2\alpha}\right]^{\frac{1}{\gamma(2\alpha-1)}}\label{phiech}.
\end{equation}
We can obtain the   values of $H$ and $\dot\phi$ from \eqrftw{Haprox}{phidotnc} respectively
for monomial potential (\ref{chaopot}) and substitute these values in \eqrf{efoldch} to
obtain the number of e-foldings $N_k$ as
\be
N_k = \frac{\phi_k^{\gamma}-\phi_{end}^{\gamma}}{\gamma}\left[\left(\frac{m^{4-n}}{12M^4}\right)^{\alpha-1}\frac{\alpha}{n M_p^{2\alpha}}(-1)^{2(\alpha-1)}\right]^{\frac{1}{2\alpha-1}}.
\ee
With  $\phi_{end}$ from  \eqrf{phiech}, we can obtain the expression
 for inflaton field $\phi_k$ when  mode $k$ leaves the horizon as
\begin{equation}
\phi_k = C_1^{1/\gamma}\left(N_k\gamma+\frac{n}{2}\right)^{\frac{1}{\gamma}}\label{phikch}.
\end{equation}
where, 
\begin{equation}
C_1 = \left\{\left(\frac{n (-M_P)^{2\alpha}}{\alpha}\right)\left(\frac{12M^4}{m^{4-n}}\right)^{\alpha-1}\right\}^{\frac{1}{2\alpha-1}}\label{c1cs}.
\end{equation}
The  first slow roll parameter $\epsilon_1$ can be expressed as a function of $N_k$
by substituting \eqrf{phikch} in \eqrf{eps1ch} as
\begin{equation}
\epsilon_1 = \frac{n}{2N_k \gamma+n}\label{eps1ch1}.
\end{equation}
Putting value of $\epsilon_1$ and $\epsilon_2$ from \eqrf{eps1ch1} and \eqrf{eps2ch2} in the 
definition of scalar spectral index $n_s$ (\ref{nsk}), we obtain 
\begin{equation}
n_s = 1-2\frac{(n+\gamma)}{2N_k\gamma+n},\label{nsch}
\end{equation}
which, on solving  for e-folds $N_k$ becomes
\begin{equation}
N_k = \frac{1}{2\gamma}\left(\frac{2(\gamma+n)}{1-n_s}-n\right)\label{nkch}.
\end{equation}
Using \eqrf{Haprox} and \eqrf{chaopot}, the value of potential at the end of inflation 
can be  obtained as
\begin{equation}
V_{end} = 3M_{P}^2 H_k^2 \left(\frac{\phi_{end}}{\phi_k}\right)^n\label{vech}.
\end{equation}
Substituting \eqrf{phiech} and \eqrf{phikch} in \eqrf{vech} we get
\begin{equation}
V_{end}= 3M_{P}^2 H_k^2\left(\frac{n}{2N_k\gamma+n}\right)^{\frac{n}{\gamma}}.\label{vech1}
\end{equation}
By substituting the value of $c_S$ from \eqrf{cs2nc} and $\epsilon_1$ from 
\eqrf{eps1ch1} in \eqrf{scalar}, we can express Hubble constant $H_K$ 
at the time when the Fourier mode $k$ leaves the inflationary horizon as 
\begin{equation}
H_k = \pi M_{P} \sqrt{8 A_S \left(\frac{n}{2N_k \gamma+n}\right)\left(\frac{1}{\sqrt{2\alpha-1}}\right)}.\label{hkch}
\end{equation}
Using  \eqrf{nkch}  we can express \eqrf{vech1} and \eqrf{hkch}  for $V_{end}$ and 
$H_K$ respectively in terms of $n_s$. Further, these expressions can be used to
 obtain the reheating temperature $T_{re}$, given by (\ref{tre1}), and  
number of e-folds during reheating $N_{re}$, given by (\ref{nre1}), as a function of 
 spectral index $n_s$.  

\begin{figure}[!h]
\centering
        \subfigure{
           
          \includegraphics[width=7cm, height=6cm]{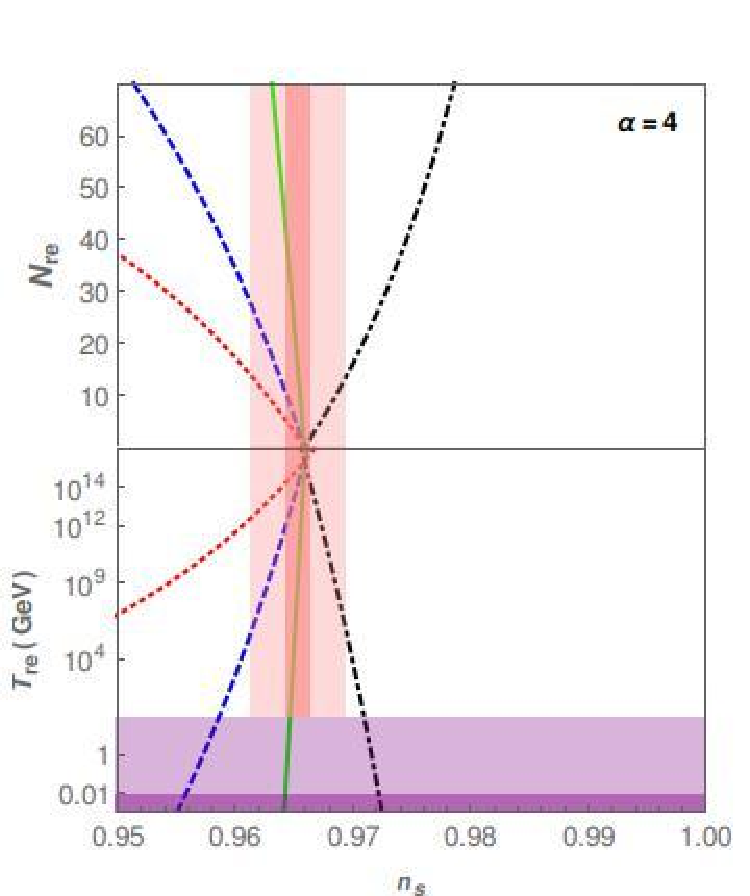}
        }
        \subfigure{
           
          \includegraphics[width=7cm, height=6cm]{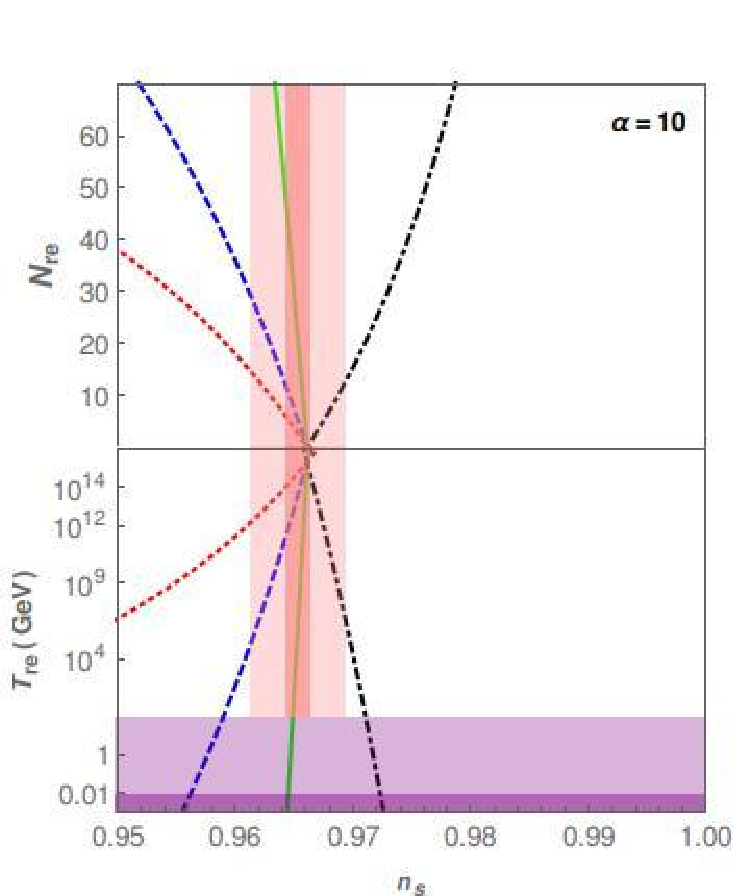}
        }\\
        \subfigure{
            
            \includegraphics[width=7cm, height=6cm]{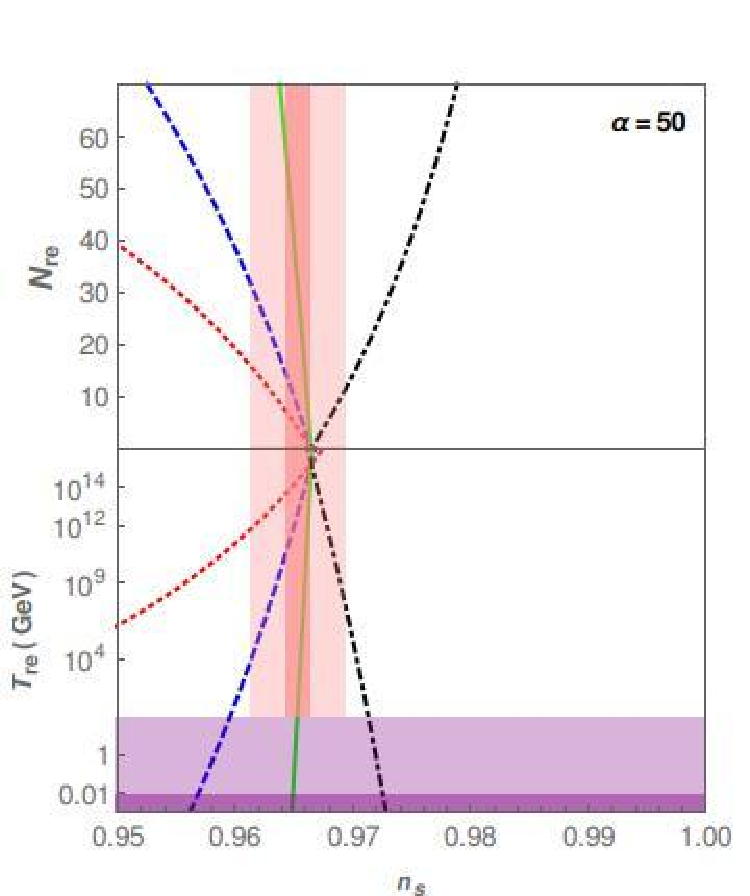}
        }
        \subfigure{
            
            \includegraphics[width=7cm, height=6cm]{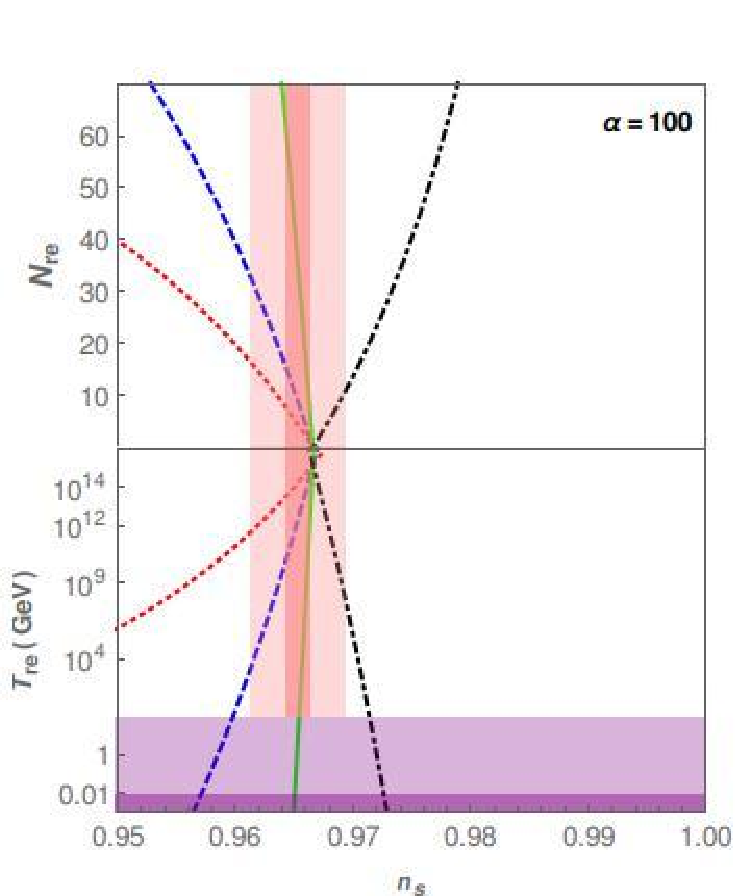}
        }

    \caption{ $N_{re}$ and $ T_{re}$ as function of $n_s$ for four different values of $\alpha$ of 
quadratic potential with power-law kinetic term. The vertical  pink region shows Planck-2018 bounds on $n_s$  
and dark pink region represents a precision of $10^{-3}$ from future observations 
\cite{Amendola:2016saw}.
The horizontal  purple region corresponds to $T_{re}$ of $10$ MeV from BBN and light
purple region corresponds to  100GeV of electroweak scale. 
Red dotted line corresponds to $w_{re} = -\frac{1}{3}$, 
blue dashed lines corresponds to $w_{re} = 0$, green solid line corresponds to  
$w_{re} = 0.25$ and  black dot-dashed line is for $w_{re}=1$.}
\label{fig:nretrephi2pl}  
 \end{figure}
\begin{figure}[!h]
     \begin{center}
        \subfigure{
           
          \includegraphics[width=7cm, height=6cm]{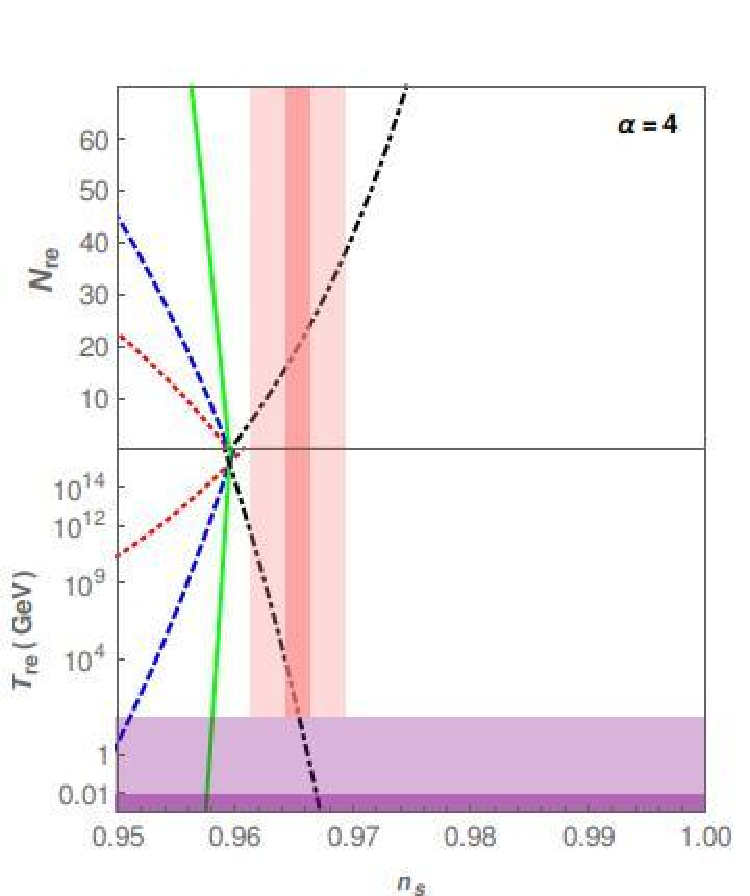}
        }
        \subfigure{
           
          \includegraphics[width=7cm, height=6cm]{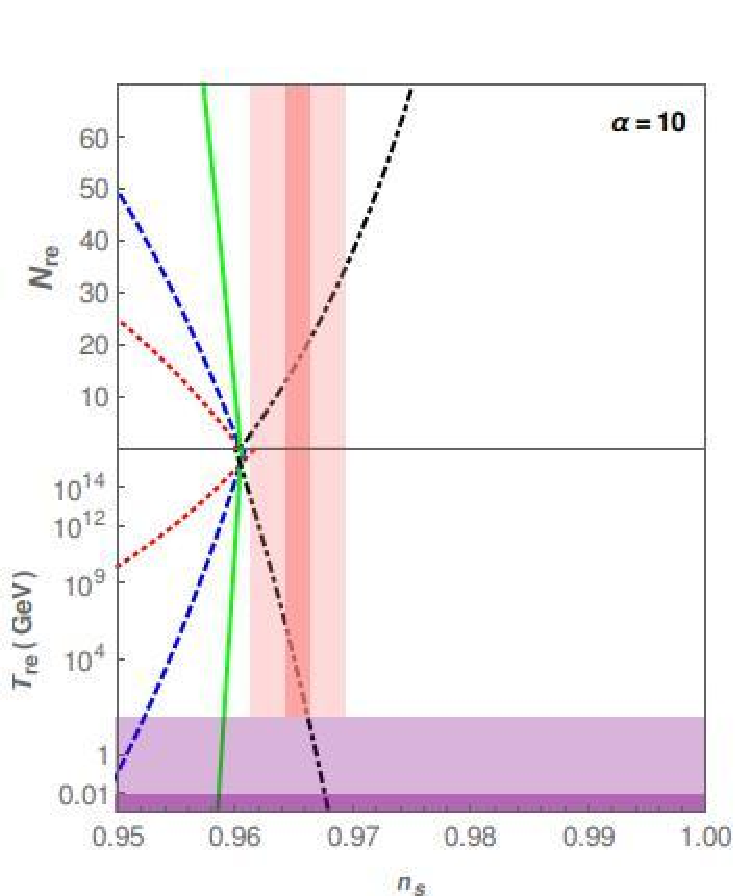}
        }\\
        \subfigure{
            
            \includegraphics[width=7cm, height=6cm]{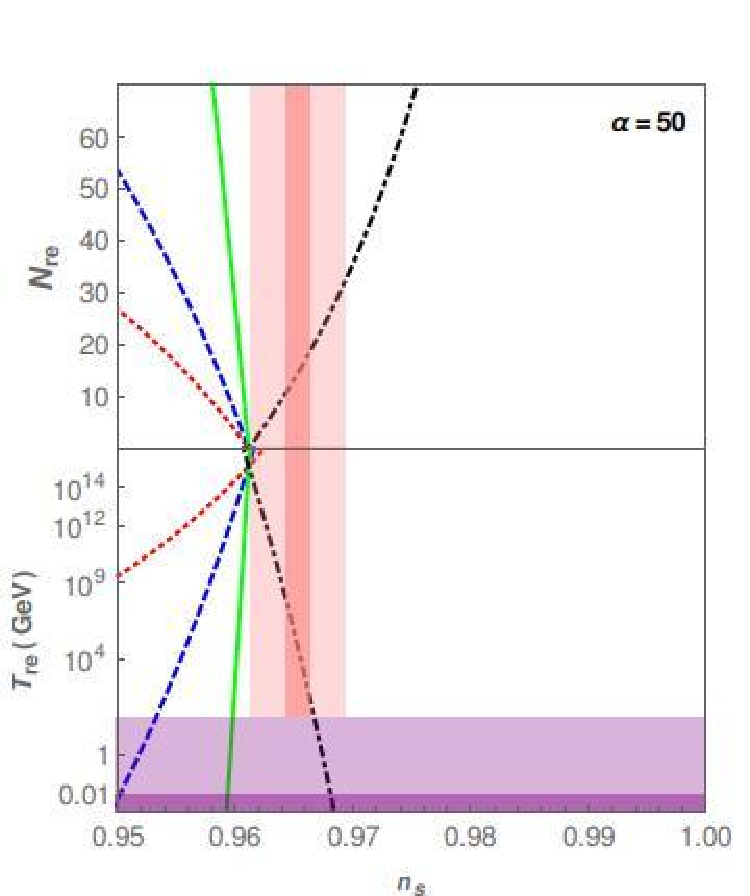}
        }
        \subfigure{
            
            \includegraphics[width=7cm, height=6cm]{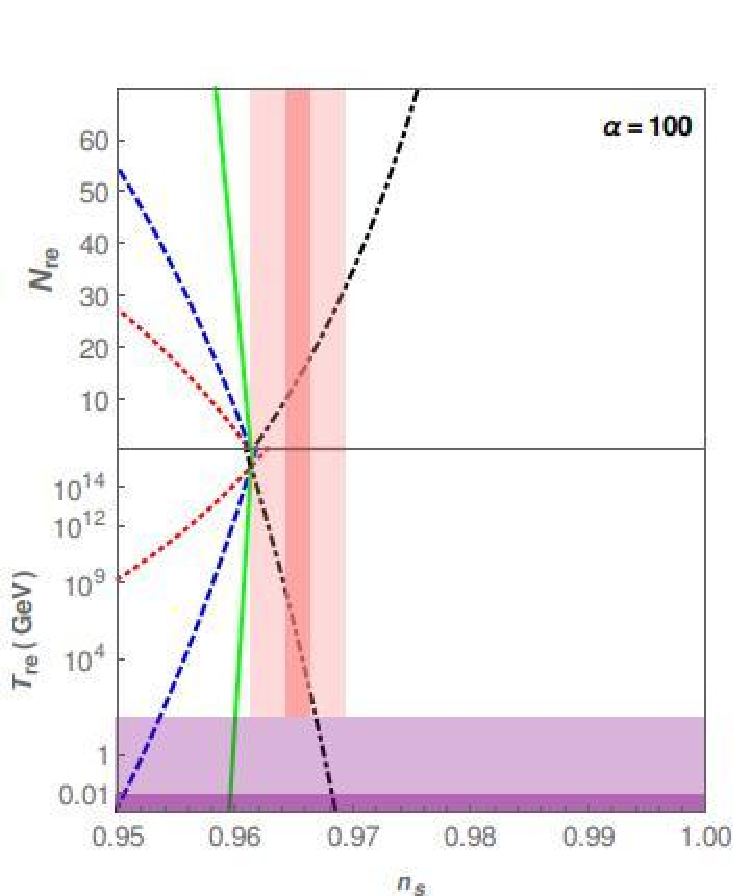}
        }
    \end{center}
\caption{$N_{re}$ and $ T_{re}$ as function of $n_s$ for four different values of $\alpha$ of 
quartic potential with power-law kinetic term. The vertical  pink region shows Planck-2018 bounds on $n_s$  
and dark pink region represents a precision of $10^{-3}$ from future observations 
\cite{Amendola:2016saw}.
The horizontal  purple region corresponds to $T_{re}$ of $10$ MeV from BBN and light
purple region corresponds to  100GeV of electroweak scale. 
Red dotted line corresponds to $w_{re} = -\frac{1}{3}$, 
blue dashed lines corresponds to $w_{re} = 0$, green solid line corresponds to  
$w_{re} = 0.25$ and  black dot-dashed line is for $w_{re}=1$.} 
\label{fig:nretrephi4pl}   
 \end{figure}
The variation of $N_{re}$ and $T_{re}$ as a function of  $n_s$, for
various values of effective equation of states, is depicted in Fig.~\ref{fig:nretrephi2pl}
along with Planck-2018 bounds  $n_s = 0.9853\pm 0.0041$. Here we choose quadratic $n=2$ and
quartic $n=4$ potentials for our analysis. It is evident from Fig.~\ref{fig:nretrephi2pl}
that, for both of these potentials, the variation of $T_{re}$ and $N_{re}$ with respect to
$n_s$ is independent of the power of kinetic term $\alpha$.
Again we imposing the bounds  on $T_{re}$, i.e., $T_{re}>100$ GeV to obtain bounds on $n_s$ for 
various equation of states $w_{re}$ by solving \eqrf{tre1}. 
Now the tensor-to-scalar ratio $r$, (\ref{rttsr}), for monomial potential with power-law
kinetic term can be obtained  using the expressions for $ c_S$, (\ref{cs2nc}) and 
$\epsilon_1$  (\ref{eps1ch1}), as  
\begin{equation}
r = \left(\frac{1}{\sqrt{2\alpha-1}}\right) \left(\frac{16n}{2N_k\gamma+n}\right).\label{rnkch}
\end{equation}

Using the bounds on $n_s$, obtained from reheating consideration, we get the bounds on
$N_k$ and tensor-to-scalar ratio $r$ for various $w_{re}$ from \eqrftw{nkch}{rnkch}.
These   bounds on $n_s$, $N_k$ and r, thus obtained, for 
quadratic and quartic potential with power-law kinetic term are provided in 
Table~\ref{tab:rnsnkpophitwo} and Table~\ref{tab:rnsnkpophifour}. The bounds on $n_s$ obtained 
from reheating are independent of $\alpha$ for  quadratic potentials. However,
the bounds obtained on tensor-to-scalar ratio $r$ depend on $\alpha$ for both the potentials.
It can be seen from Table~\ref{tab:rnsnkpophifour} that, with $\alpha=4$, the  bounds
on tensor-to-scalar ratio $0.086\ge r \ge 0.0740$ lie slightly above than the joint
BICEP2/Keck array and Planck-2018 bound $r<0.06$ \cite{Ade:2018gkx} for physically plausible 
range $0 \le w_{re} \le 0.25$ for effective equation of state during reheating. But, for 
larger values of $\alpha$ the bounds on $r$ are in agreement with BICEP2/Keck array bound.

\begin{table}[h]
\begin{tabular}{ |c|c|c|c|c| } 
\hline
$\alpha $ &  Equation of state  & $ n_s $ & $ N_k $ & r \\
\hline
\multirow{3}{4em}{$\alpha=4$}
& $-1/3\leq{w_{re}}\leq{0} $& ${0.9273}\leq{n_s}\leq{0.9586}$&$27.02\leq{N_k}\leq{47.85}$  &$ {0.1098}\geq{r}\geq{0.0625} $ \\ 
& ${0}\leq{w_{re}}\leq{0.25}$ &  ${0.9586}\leq{n_s}\leq{0.9640}$&$47.85\leq{N_k}\leq{56.07}$ & ${0.0625}\geq{r}\geq{0.0534}$  \\
&${0.25}\leq{w_{re}}\leq{1} $& $ {0.9640}\leq{n_s}\leq{0.9709}$ &$56.07\leq{N_k}\leq{68.33}$&$ {0.0534}\geq{r}\geq{0.0439}$  \\ 
\hline
\multirow{3}{4em}{$\alpha=10$}
& $-1/3\leq{w_{re}}\leq{0} $& ${0.9287}\leq{n_s}\leq{0.9589}$&$27.53\leq{N_k}\leq{48.27}$ &$ {0.0655}\geq{r}\geq{0.0376} $  \\ 
&  ${0}\leq{w_{re}}\leq{0.25}$ &  ${0.9589}\leq{n_s}\leq{0.9649}$ &$48.27\leq{N_k}\leq{56.45}$  & ${0.0376}\geq{r}\geq{0.0322}$ \\
&${0.25}\leq{w_{re}}\leq{1} $& $ {0.9649}\leq{n_s}\leq{0.9710}$ &$56.45\leq{N_k}\leq{68.67}$ &$ {0.0322}\geq{r}\geq{0.0265}$\\ 
\hline
\multirow{3}{4em}{$\alpha=50$} 
&  $-1/3\leq{w_{re}}\leq{0} $& ${0.9307}\leq{n_s}\leq{0.9596}$ &$28.38\leq{N_k}\leq{48.96}$ &$ {0.0278}\geq{r}\geq{0.0163} $ \\
& ${0}\leq{w_{re}}\leq{0.25}$ &  ${0.9596}\leq{n_s}\leq{0.9653}$ &$48.96\leq{N_k}\leq{57.09}$& ${0.0163}\geq{r}\geq{0.0140}$  \\ 
& ${0.25}\leq{w_{re}}\leq{1} $& $ {0.9653}\leq{n_s}\leq{0.9713}$ &$57.09\leq{N_k}\leq{69.22}$&$ {0.0140}\geq{r}\geq{0.0115}$  \\ 
\hline
\multirow{3}{4em}{$\alpha=100$} 
& $-1/3\leq{w_{re}}\leq{0} $& ${0.9315}\leq{n_s}\leq{0.9598}$&$28.73\leq{N_k}\leq{49.25}$ &$ {0.0194}\geq{r}\geq{0.0114} $  \\ 
& ${0}\leq{w_{re}}\leq{0.25}$ &  ${0.9598}\leq{n_s}\leq{0.9654}$ &$49.25\leq{N_k}\leq{57.36}$& ${0.0114}\geq{r}\geq{0.0098}$  \\ 
& ${0.25}\leq{w_{re}}\leq{1} $& $ {0.9654}\leq{n_s}\leq{0.9714}$&$57.36\leq{N_k}\leq{69.45}$ &$ {0.0098}\geq{r}\geq{0.0081}$  \\  
\hline
\end{tabular}
\caption{The allowed values of spectral index $n_s$ and number of e-folds $N_k$ for various values of
$\alpha$ for quadratic potential with power-law kinetic term  considering $T_{re}\geq{100GeV}$.}
\label{tab:rnsnkpophitwo}
\end{table} 
\begin{table}[h]
\begin{tabular}{ |c|c|c|c|c| } 
\hline
$\alpha $ &  Equation of state  & $ n_s $  & $ N_k $ & r\\
\hline
\multirow{3}{4em}{$\alpha = 4 $}
& $-1/3\leq{w_{re}}\leq{0} $& ${0.9152}\leq{n_s}\leq{0.9510}$ &$27.62\leq{N_k}\leq{48.36}$ &$ {0.1495}\geq{r}\geq{0.0863} $ \\ 
& ${0}\leq{w_{re}}\leq{0.25}$ &  ${0.9510}\leq{n_s}\leq{0.9581}$ &$48.36\leq{N_k}\leq{56.53}$& ${0.0863}\geq{r}\geq{0.0740}$  \\
&${0.25}\leq{w_{re}}\leq{1} $& $ {0.9581}\leq{n_s}\leq{0.9654}$ &$56.53\leq{N_k}\leq{68.69}$ &$ {0.0740}\geq{r}\geq{0.0610}$ \\ 
\hline
\multirow{3}{4em}{$\alpha=10$}
& $-1/3\leq{w_{re}}\leq{0} $& ${0.9180}\leq{n_s}\leq{0.9522}$ &$28.06\leq{N_k}\leq{48.73}$&$ {0.0867}\geq{r}\geq{0.0505} $  \\ 
&  ${0}\leq{w_{re}}\leq{0.25}$ &  ${0.9522}\leq{n_s}\leq{0.9590}$ &$48.73\leq{N_k}\leq{56.87}$& ${0.0505}\geq{r}\geq{0.0433}$  \\
&${0.25}\leq{w_{re}}\leq{1} $& $ {0.9590}\leq{n_s}\leq{0.9662}$ &$56.87\leq{N_k}\leq{68.99}$ &$ {0.0433}\geq{r}\geq{0.0357}$\\ 
\hline
\multirow{3}{4em}{$\alpha=50$} 
&  $-1/3\leq{w_{re}}\leq{0} $& ${0.9209}\leq{n_s}\leq{0.9533}$ &$28.88\leq{N_k}\leq{49.10}$&$ {0.0364}\geq{r}\geq{0.0215} $ \\
& ${0}\leq{w_{re}}\leq{0.25}$ &  ${0.9533}\leq{n_s}\leq{0.9598}$ & $49.10\leq{N_k}\leq{57.48}$& ${0.0215}\geq{r}\geq{0.0185}$  \\ 
& ${0.25}\leq{w_{re}}\leq{1} $& $ {0.9598}\leq{n_s}\leq{0.9667}$ & $57.48\leq{N_k}\leq{69.52}$ &$ {0.0185}\geq{r}\geq{0.0153}$  \\ 
\hline
\multirow{3}{4em}{$\alpha=100$} 
& $-1/3\leq{w_{re}}\leq{0} $& ${0.9219}\leq{n_s}\leq{0.9536}$  & $29.23\leq{N_k}\leq{49.69}$ &$ {0.0253}\geq{r}\geq{0.0150} $ \\ 
& ${0}\leq{w_{re}}\leq{0.25}$ &  ${0.9536}\leq{n_s}\leq{0.9600}$  & $49.69\leq{N_k}\leq{57.75}$ & ${0.0150}\geq{r}\geq{0.0129}$\\ 
& ${0.25}\leq{w_{re}}\leq{1} $& $ {0.9600}\leq{n_s}\leq{0.9668}$ & $57.75\leq{N_k}\leq{69.76}$ &$ {0.0129}\geq{r}\geq{0.0108}$ \\  
\hline
\end{tabular}
\caption{The allowed values of spectral index $n_s$ and number of e-folds $N_k$ for various values of
$\alpha$ for quartic potential with power-law kinetic term considering $T_{re}\geq{100GeV}$.}
\label{tab:rnsnkpophifour}
\end{table}

\begin{figure}[h]
\centering
\subfigure[ $N_k$ vs $n_s$  for  quadratic potential]{
         \includegraphics[width=6cm, height=5cm]{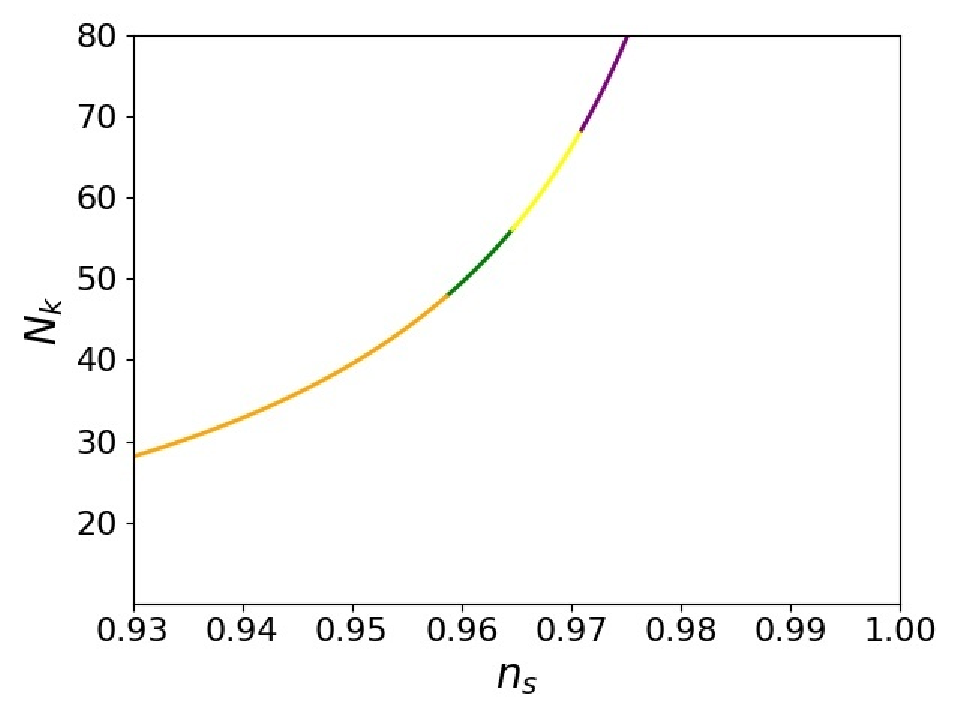}
          }
\subfigure[$N_k$ vs $n_s$  for   quartic potential]{
         \includegraphics[width=6cm, height=5cm]{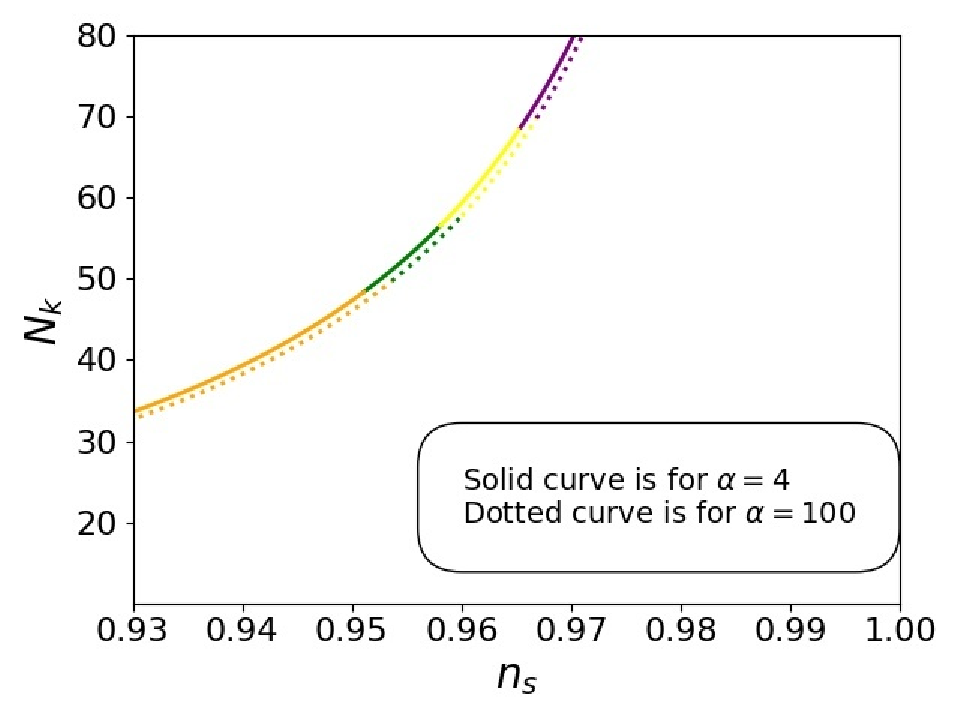}
        }
       
\caption{ $N_k$ as function of $n_s$ for quadratic potential and quartic potential 
with power law kinetic term. }
\label{fig:nknsphi2pl}
\end{figure}

\begin{figure}
\centering
\subfigure{
\includegraphics[width=7cm,height=6cm]{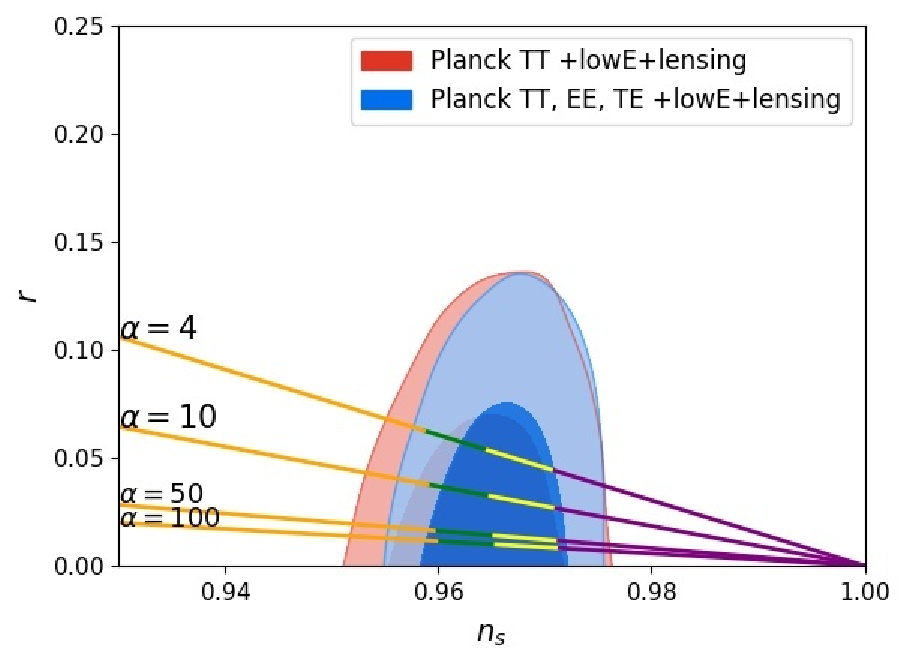}
 }
 \subfigure{
\includegraphics[width=7cm,height=6cm]{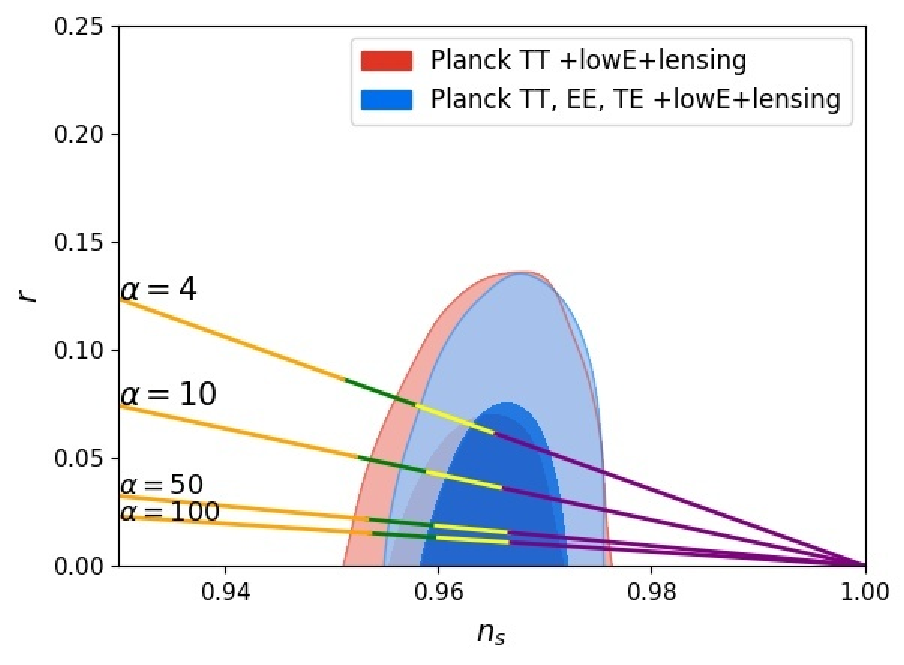}
}
\caption{r vs $n_s$ predictions for quadratic and quartic potentials with  four different choice of $\alpha$  along with joint 68\%C.L. and 95\%C.L. Planck-2018 constraints. Here the orange region corresponds to $w_{re}\le 0$, green region corresponds to 
$0\le w_{re}\le 0.25$, yellow region shows $0.25\le w_{re}\le 1$ and purple region corresponds 
to $w_{re}>1$.}
\label{fig:rnsphitwoandfour}
\end{figure}

Plots for $N_k$ vs $n_s$ for quadratic and quartic potentials are shown in 
Fig.~\ref{fig:nknsphi2pl}. Here we have chosen only one value $\alpha=4$ for
quadratic potential, as the variation of 
$N_k$ with respect to $n_s$ is independent of $\alpha$. 
In case of quartic potential also we have chosen only the smallest and largest values
of $\alpha$, because the variation of functional dependence of $N_k$ on $n_s$ with respect 
to $\alpha$ is very small. Fig.~\ref{fig:rnsphitwoandfour} depicts the $r$ vs $n_s$ predictions
for quadratic and quartic potential for  different values of $\alpha$ and $w_{re}$, along with
 joint $68\%$ and $95\%$ C.L. constraints from Planck-2018.
 It can be seen from Fig.~\ref{fig:rnsphitwoandfour}  that $r$ vs $n_s$ predictions
for the quadratic  potential with power-law kinetic term lie within 68\% C.L.  of Planck-2018
constraints for physically plausible range of $0\le w_{re}\le 0.25$.
However, for quartic potential  the equation of state during reheating should be greater than
 0.25 for $r$-$n_s$ predictions to lie within 68\% C.L.  of Planck-2018
constraints.

\subsection{Exponential potential}
We  now consider the exponential potential with power-law kinetic term.
This potential have the  following form
\begin{equation}
V(\phi) = V_0 \exp\left(-\sqrt{\frac{2}{q}}\frac{\phi}{M_{P}}\right).\label{exppot}
\end{equation}
 In case of inflation with canonical scalar field this potential provides 
power-law expansion, $a(t)\propto t^q$ , 
for flat universe \cite{Lucchin:1984yf,Halliwell:1986ja,Yokoyama:1987an}. The power-law
solutions can also be obtained with this potential in DBI framework \cite{Chimento:2010un}.

We can obtain the slow-roll parameters $\epsilon_1$ and $\epsilon_2$ for this potential 
using \eqrf{eps1nc} and \eqrf{eps2nc} as
\bea
\epsilon_1 &=& \left[\frac{1}{\alpha} \left(\frac{3M^4}{V_0}\right)^{\alpha-1} \left(\frac{1}{\sqrt{q}}\right)^{2\alpha}\frac{1}{\exp\left(-\sqrt{\frac{2}{q}}\frac{\phi(\alpha-1)}{M_{P}}\right)}\right]^{\frac{1}{2\alpha-1}}\label{eps1exp},\\
\epsilon_2 &=& 2\epsilon_1 \left(\frac{\alpha-1}{2\alpha-1}\right)\label{eps2exp}.
\eea
Now we evaluate $\phi_{end}$, the value of inflaton field at the end of inflation, 
by setting $\epsilon_1=1$ as
\begin{equation}
\phi_{end} = -\frac{M_{P}}{\alpha-1}\sqrt{\frac{q}{2}}\ln\left[\frac{1}{\alpha}\left(\frac{3M^4}{V_0}\right)^{\alpha-1}\left(\sqrt{\frac{1}{q}}\right)^{2\alpha}\right].\label{expphie}
\end{equation}
To obtain the number of e-foldings $N_k$ from the time when the Fourier
mode $k$ leaves the Hubble radius to the end of inflation, for (\ref{exppot}), we put values of 
$H$ and $\dot\phi$ from \eqrftw{Haprox}{phidotnc} into  \eqrf{efoldch}, and on integrating
it we get
\be
N_k = \frac{\phi_k^{\frac{\alpha-1}{2\alpha-1}}-\phi_{end}^{\frac{\alpha-1}{2\alpha-1}}}{2^{\frac{\alpha-1}{2\alpha-1}}}\left(\frac{V_0}{3M^4}\right)^{\frac{\alpha-1}{2\alpha-1}} \left({\sqrt{\frac{q}{2}}}\right)^{\frac{2\alpha}{2\alpha-1}}\alpha^{\frac{1}{2\alpha-1}}.
\ee
Substituting  $\phi_{end}$ from \eqrf{expphie} and solving for $\phi_k$, the value of
inflaton field at horizon crossing, 
we obtain 
\begin{equation}
\phi_k = -\sqrt{\frac{q}{2}}\frac{M_{P}}{(\alpha-1)}\ln\left[\frac{1}{\alpha}\left(\frac{3M^4}{V_0}\right)^{\alpha-1}\left(\sqrt{\frac{2}{q}}\right)^{2\alpha}\left(\frac{1}{2^{\frac{\alpha}{2\alpha-1}}}+N_k \left(\frac{\alpha-1}{2\alpha-1}\right)2^{\frac{\alpha-1}{2\alpha-1}}\right)^{2\alpha-1}\right].\label{expphik}
\end{equation}
Substituting \eqrf{expphik} in \eqrf{eps1exp}, we can evaluate the first slow-roll
parameter $\epsilon_1$ at $\phi=\phi_k$ as
\begin{equation}
\epsilon_1 = \frac{2\alpha-1}{(2\alpha-1)+ 2N_k(\alpha-1)}.\label{eps1Nk}
\end{equation}
Putting \eqrf{eps1Nk} and \eqrf{eps2exp} in \eqrf{nsk}, we get the expression for spectral 
index
\begin{equation}
n_s = 1-2\frac{(3\alpha-2)}{(2\alpha-1)+2N_k(\alpha-1)}\label{nsnkexp}.
\end{equation}
Using this equation the number of e-folds $N_k$ can be expressed in terms of spectral index $n_s$ 
as
\begin{equation}
N_k = \frac{(3\alpha-2)}{(\alpha-1)(1-n_s)} - \frac{(2\alpha-1)}{2(\alpha-1)}\label{nkexp}.
\end{equation}
The value of  the potential at the end of inflation can be expressed in terms of
$H_k$ using \eqrftw{Haprox}{exppot} as
\begin{equation}
V_{end} = 3 M_{P}^2 H_k^2 \left[\frac{\exp\left(-\sqrt{\frac{2}{q}}\frac{\phi_{end}}{M_{P}}\right)}{\exp\left(-\sqrt{\frac{2}{q}}\frac{\phi_k}{M_{P}}\right)}\right].\label{vexp}
\end{equation}
Solving above equation with \eqrf{expphie} and \eqrf{expphik}
\begin{equation}
V_{end} = 3 M_{P}^2 H_k^2 \left[\frac{2\alpha-1}{(2\alpha-1)+2N_k(\alpha-1)}\right]^{\frac{2\alpha-1}{\alpha-1}}.
\end{equation}

\begin{figure}[!h]
\centering
        \subfigure{
           
          \includegraphics[width=6cm, height=5cm]{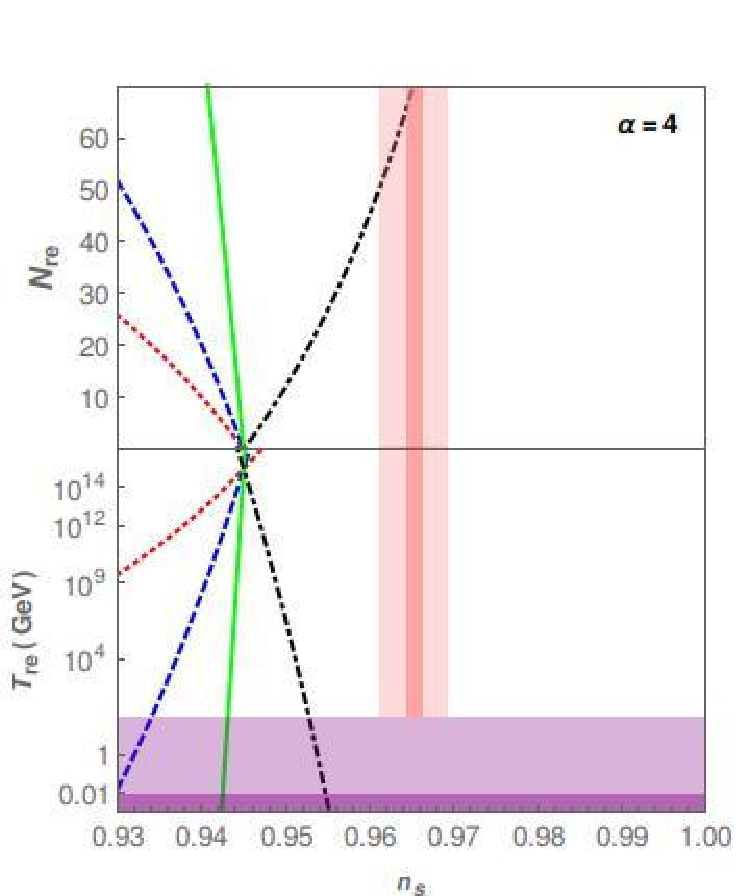}
        }
        \subfigure{
           
          \includegraphics[width=6cm, height=5cm]{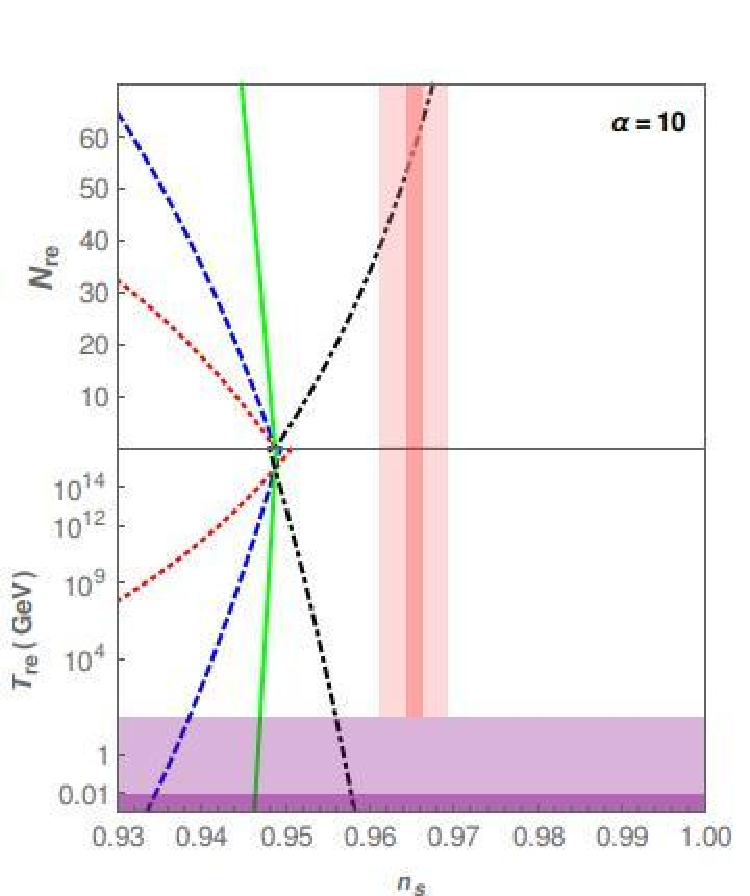}
        }
        \subfigure{
            
            \includegraphics[width=6cm, height=5cm]{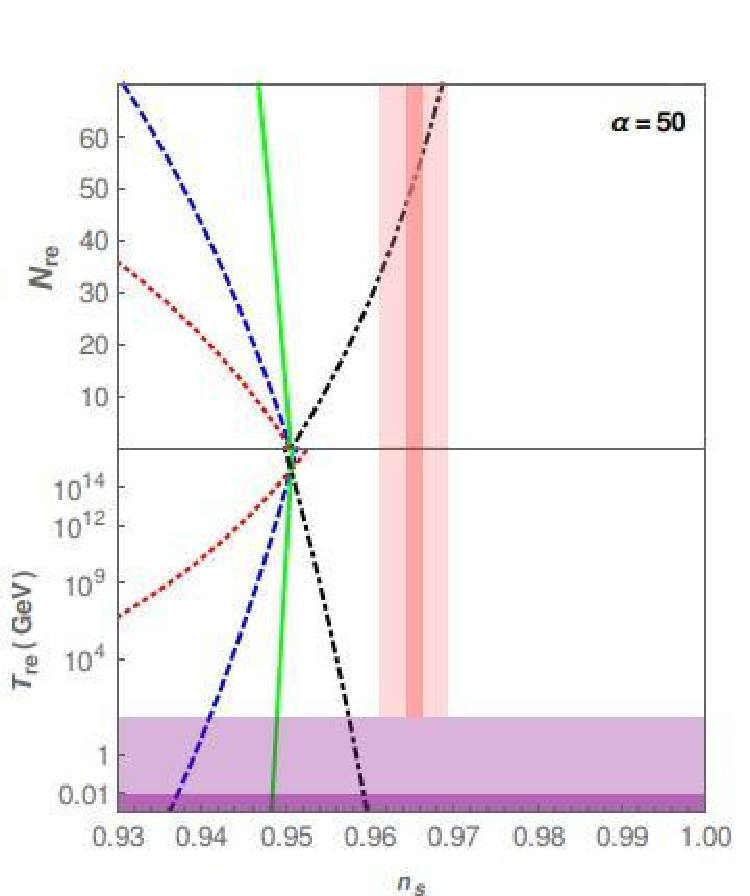}
        }
        \subfigure{
            
            \includegraphics[width=6cm, height=5cm]{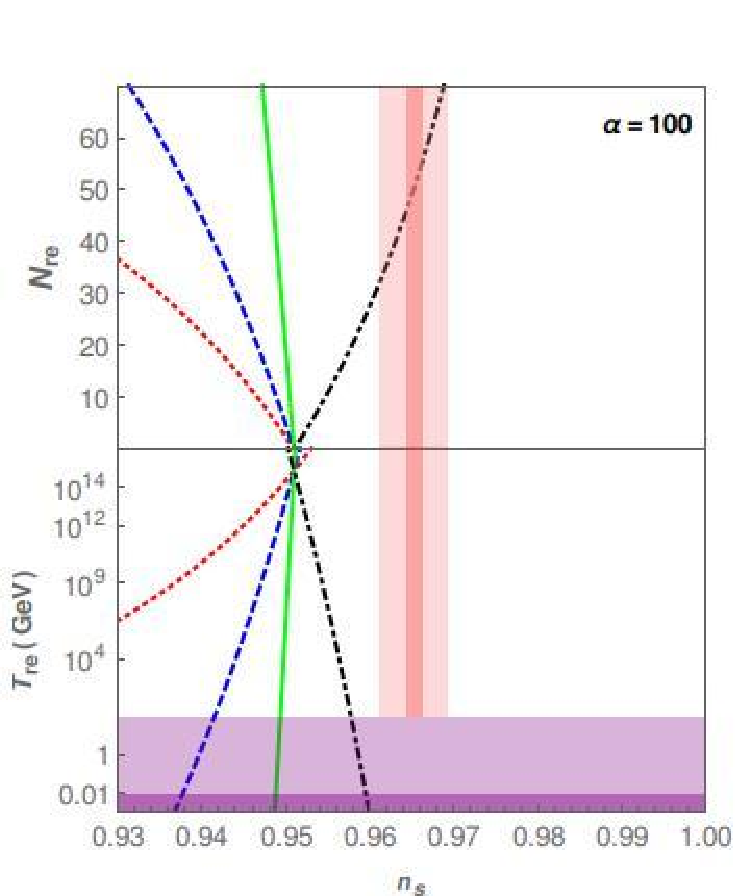}
        }
\caption{ $N_{re}$ and $ T_{re}$ as function of $n_s$ for four different values of $\alpha$ of 
exponential potential with power-law kinetic term. The vertical  pink region shows Planck-2018 bounds on $n_s$  
and dark pink region represents a precision of $10^{-3}$ from future observations 
\cite{Amendola:2016saw}.
The horizontal  purple region corresponds to $T_{re}$ of $10$ MeV from BBN and light
purple region corresponds to  100GeV of electroweak scale. 
Red dotted line corresponds to $w_{re} = -\frac{1}{3}$, 
blue dashed lines corresponds to $w_{re} = 0$, green solid line corresponds to  
$w_{re} = 0.25$ and  black dot-dashed line is for $w_{re}=1$.}
\label{fig:nretreexp} 
\end{figure}

The Hubble constant at $\phi = \phi_k$,
can be obtained by substituting the expression for speed of sound $c_S$, (\ref{cs2nc}),
and slow-roll parameter $\epsilon_1$,(\ref{eps1Nk}) in \eqrf{scalar} as
\begin{equation}
H_k = \pi M_{P}\sqrt{8 A_S  \frac{2\alpha-1}{(2\alpha-1)+ 2N_k(\alpha-1)}\frac{1}{\sqrt{2\alpha-1}}}\label{Hkexp}.
\end{equation}

Using equation \eqrf{Hkexp},  \eqrf{vexp}, and \eqrf{nkexp}, we can obtain  $H_k$ and $V_{end}$
as a function of $n_s$. Again, by using these expressions for $H_k$ and $V_{end}$,
the reheating temperature $T_{re}$ and the number of e-folds during reheating $N_{re}$ can
be obtained in terms of $n_s$ from \eqrftw{tre1}{nre1} respectively.
The variation of $N_{re}$ and $T_{re}$ with respect to $n_s$ for various choices of $\alpha$ and
effective equation of state during reheating is shown in
Fig.~\ref{fig:nretreexp}.

By demanding $T_{re}>100$ GeV we obtain bounds on $n_s$ using \eqrf{tre1} for various
values of $w_{re}$. Again from these bounds on $n_s$, the bounds on $N_k$ can be
obtained using \eqrf{nkexp}. The tensor-to-scalar ratio $r$ for exponential potential
with power-law kinetic term can be obtained by substituting \eqrf{cs2nc}, \eqrf{eps1Nk}  
in \eqrf{rttsr} as
\begin{equation}
r = \frac{16\sqrt{2\alpha-1}}{2\alpha-1+2N_k(\alpha-1)}, \quad \alpha > 1\label{rexp}.
\end{equation}

 Using this expression we can get bounds on $r$ from the bounds on $N_k$, obtained by 
reheating consideration. These bounds   on $n_s$, $N_k$ and $r$ for exponential potential are
 listed in  Table~\ref{table:rnsnkexp}. It can be seen from the Table that, 
with $\alpha=4$, the bounds $r$, i.e., $0.139\ge r\ge 0.12$ are higher than the joint
BICEP2/Keck array and Planck-2018 bounds  $r<0.06$ \cite{Ade:2018gkx} for physically plausible
range $0\le w_{re}\le 0.25$. However, for this range of $w_{re}$, the bounds on $r$ are
compatible with joint BICEP2/Keck array and Planck-2018 bounds for larger values of $\alpha$.

\begin{table}[!hbt]
\begin{tabular}{ |c|c|c|c|c| } 
\hline
$\alpha $ &  Equation of state  & $ n_s $ & $N_k$ & $ r $ \\
\hline
\multirow{3}{4em}{$\alpha=4 $}
& $-1/3\leq{w_{re}}\leq{0} $ & $ {0.8888}\leq{n_s}\leq{0.9341}$&$31.81\leq{N_k}\leq{54.45}$ &$ {0.2353}\geq{r}\geq{0.1395} $  \\ 
& ${0}\leq{w_{re}}\leq{0.25}$ &  ${0.9341}\leq{n_s}\leq{0.9431}$ &$54.45\leq{N_k}\leq{63.30}$ & ${0.1395}\geq{r}\geq{0.1204}$ \\
&${0.25}\leq{w_{re}}\leq{1} $& $ {0.9431}\leq{n_s}\leq{0.9528}$&$63.30\leq{N_k}\leq{76.44}$ &$ {0.1204}\geq{r}\geq{0.0999}$  \\ 
\hline
\multirow{3}{4em}{$\alpha=10$}
& $-1/3\leq{w_{re}}\leq{0} $& ${0.8967}\leq{n_s}\leq{0.9385}$ &$30.1371\leq{N_k}\leq{51.39}$ &$ {0.1286}\geq{r}\geq{0.0765} $ \\ 
& ${0}\leq{w_{re}}\leq{0.25}$ &  ${0.9385}\leq{n_s}\leq{0.9469}$ &$51.39\leq{N_k}\leq{59.72}$ & ${0.0765}\geq{r}\geq{0.0660}$ \\
& ${0.25}\leq{w_{re}}\leq{1} $& $ {0.9469}\leq{n_s}\leq{0.9559}$ &$59.72\leq{N_k}\leq{72.09}$&$ {0.0660}\geq{r}\geq{0.0549}$ \\ 
\hline
\multirow{3}{4em}{$\alpha=50$} 
& $-1/3\leq{w_{re}}\leq{0} $& ${0.9020}\leq{n_s}\leq{0.9410}$ &$30.01\leq{N_k}\leq{50.53}$&$ {0.0527}\geq{r}\geq{0.0317} $ \\
& ${0}\leq{w_{re}}\leq{0.25}$ &  ${0.9410}\leq{n_s}\leq{0.9489}$ &$50.53\leq{N_k}\leq{58.58}$& ${0.0317}\geq{r}\geq{0.0274}$  \\ 
& ${0.25}\leq{w_{re}}\leq{1} $& $ {0.9489}\leq{n_s}\leq{0.9575}$ &$58.58\leq{N_k}\leq{70.54}$&$ {0.0274}\geq{r}\geq{0.0229}$  \\ 
\hline
\multirow{3}{4em}{$\alpha=100$} 
& $ -1/3\leq{w_{re}}\leq{0} $ & ${0.9034}\leq{n_s}\leq{0.9415}$& $ 30.25\leq{N_k}\leq{50.64} $ & $ {0.0366}\geq{r}\geq{0.0221} $  \\ 
& $ {0}\leq{w_{re}}\leq{0.25} $ &  ${0.9415}\leq{n_s}\leq{0.9493} $& $ 50.64\leq{N_k}\leq{58.63} $  & ${0.0221}\geq{r}\geq{0.0192} $ \\ 
& $ {0.25}\leq{w_{re}}\leq{1} $ & $ {0.9493}\leq{n_s}\leq{0.9578}$ & $ 58.63\leq{N_k}\leq{70.52} $  & $ {0.0192}\geq{r}\geq{0.0159} $ \\  
\hline
\end{tabular}
\caption{The allowed values of spectral index $n_s$ and number of e-folds $N_k$ for various values of
 $\alpha$ for exponential potential with power-law kinetic term, considering $T_{re}\geq{100GeV}$.}
\label{table:rnsnkexp}
\end{table}

\begin{figure}[!htp]
\centering
\subfigure[ $N_k$ vs $n_s$  for  exponential potential for $\alpha = 4, 10, 100$]{
         \includegraphics[width=7cm, height=5.5cm]{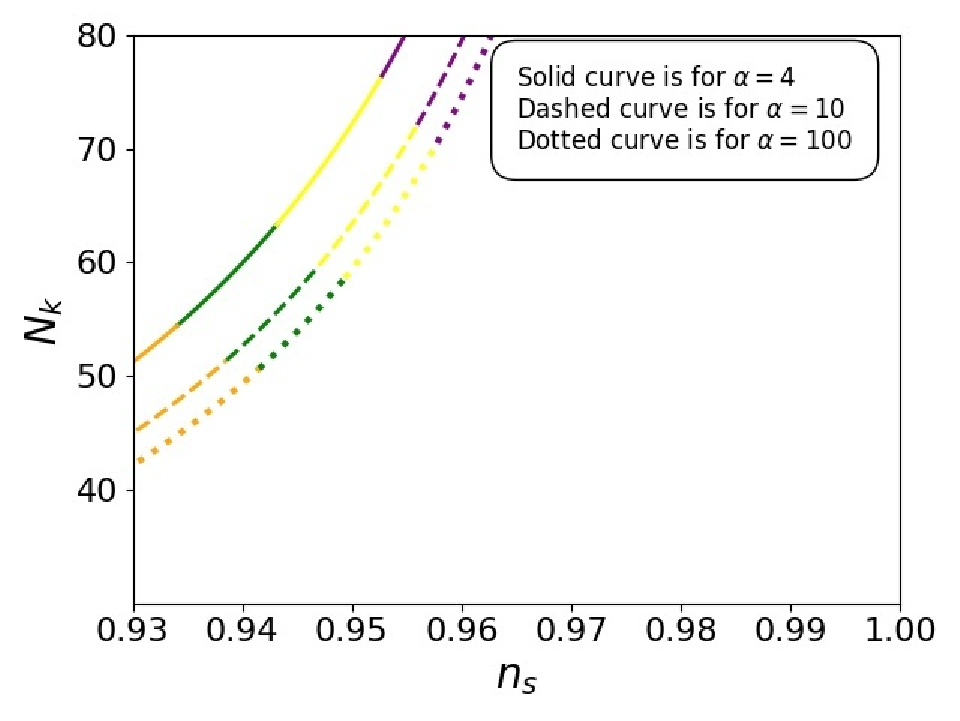}
          }
\subfigure[$r$ vs $n_s$  for for exponential with $\alpha = 4, 10, 50, 100$]{
         \includegraphics[width=7cm, height=5.5cm]{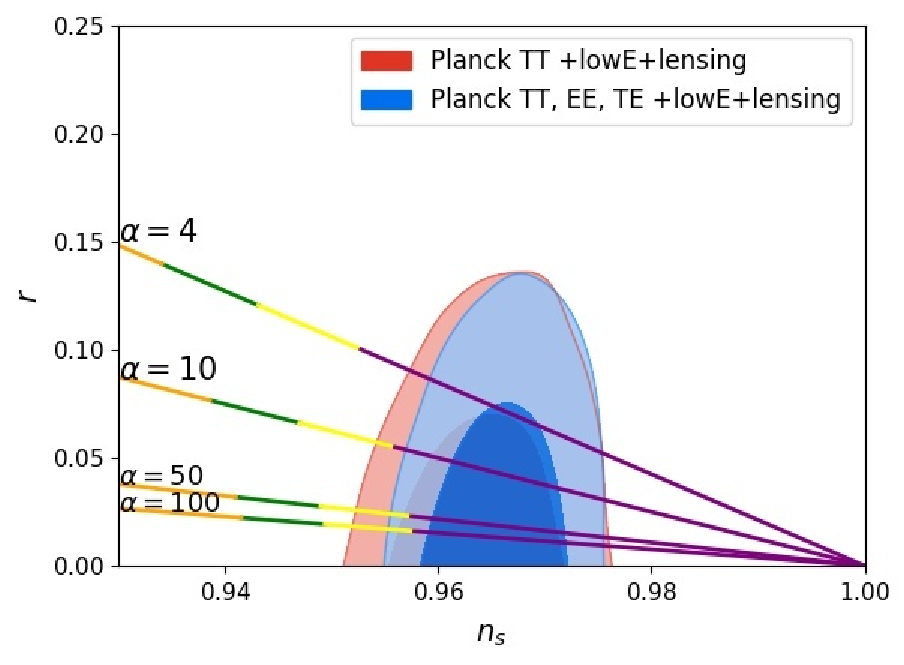}
}
\caption{ In the left panel $N_k$ as function of $n_s$ is shown for  $\alpha = 4, 50$ and, 100 of exponential potential with power law kinetic term. In the right panel predictions of r vs $n_s$ for exponential potential with power law kinetic term
along with joint 68\%CL and 95\%CL Planck-2018 constraints is shown  for four choices of $\alpha$ ($\alpha = 4, 10, 50$ and, $100$).
Here in both panels the orange region corresponds to $w_{re}\le 0$, green region corresponds to 
$0\le w_{re}\le 0.25$, yellow region shows $0.25\le w_{re}\le 1$ and purple region corresponds 
to $w_{re}>1$.}
\label{fig:rnsexp}
\end{figure}
The plots between $N_k$ and $n_s$ are shown in left panel of Fig.~\ref{fig:rnsexp} for various 
values of $\alpha$ and $w_{re}$. 
The $r-n_s$ predictions, along with joint $68\%$ and $95\%$ C.L. Planck-2018 constraints
 constraints, for this case are shown  in the right panel of Fig.~\ref{fig:rnsexp}. 
It can bee seen from the figure that, for all values of $\alpha$, the effective 
equation of state during
reheating $w_{re}$ should be greater than $1$ to satisfy Planck-2018 joint constraints on 
$r$ and $n_s$, which violates causality. 
 
\section{Conclusion}\label{conclusion}
K-inflation \cite{ArmendarizPicon:1999rj,Garriga:1999vw} is an alternative to the standard 
single field slow-roll inflation. In this case the noncanonical kinetic term of 
the scalar field drives inflation. 
This scenario has an advantage over the canonical single field inflation as it
increases the viability of various inflaton potentials, ruled out from Planck CMB
observations, by reducing the tensor-to-scalar ratio. In this work we analyze models of
K-inflation in the light of reheating.
The phase of reheating can be parameterized in terms of three parameters, namely reheating
temperature $T_{re}$, effective equation of state of cosmic fluid during reheating $w_{re}$ and
number of e-folds during reheating $N_{re}$. These three parameters can be related to the
amplitude of scalar perturbations, spectral index and other inflationary parameters depending
on inflaton kinetic term and potential and can be used to constrain models of inflation
(see \cite{Dai:2014jja,Munoz:2014eqa,Cook:2015vqa} for constraints on canonical single field
inflation).
We  derive expressions for $T_{re}$ and $N_{re}$ in terms of $w_{re}$, $n_s$ and other
inflationary parameters, and then use these expressions to constrain  models of K-inflation
having  kinetic term of DBI form and power-law form. With DBI kinetic term we choose monomial
and natural inflation potential and with power-law kinetic term we choose monomial and exponential
potential. In \cite{Podolsky:2005bw} it was shown that the  equation of state during reheating
$w_{re}$ should lie between $0$ to $0.25$ for various reheating scenario. By imposing
$0\le w_{re}\le 0.25$ and demanding that the reheating temperature $T_{re}>100$ GeV for weak scale
dark matter production, we find bounds on $n_s$ and number of e-foldings $N_k$ from the time
when the mode $k$ corresponding to the pivot scale, $k_0=0.05$
M\textsubscript{pc}\textsuperscript{-1} leaves the Hubble radius during inflation to
the end of inflation.  These bounds on $n_s$ and $N_k$ can be transferred the bounds on
tensor-to-scalar ratio $r$, and hence the allowed region in $n_s-r$ plane for models of inflation
is restricted.

The bounds obtained for $N_k$ and $r$ for  K-inflation with DBI kinetic term and 
monomial potentials $V \sim \phi^n$ are shown in Table:~\ref{table:nsnkrpoldbi} and 
the $r-n_s$ predictions for various equation of state during reheating are 
shown in Fig.~\ref{fig:nkrnspoldbi}.  We find that
the tensor-to-scalar ratio $r>0.0786$ for $w_{re}\le 1$ in case  of quartic potential, which 
is greater than the joint BICEP2/Keck array and  Planck-2018 bound $r<0.06$ \cite{Ade:2018gkx}. 
The $r-n_s$ predictions for $n=2/3$ and $n=1$
lie within the Planck-2018 $1\sigma$ constraints for $w_{re}< 0$. The bounds on $N_k$ 
and $r$ 
for natural inflation potential are shown in Table.~\ref{table:nsnkrnatdbi} and the 
predictions for $r-n_s$  are
represented in Fig.~\ref{fig:nkrnsnatdbi}. We find that the  natural inflation with DBI 
kinetic term is
compatible with Planck-2018 observations for physically plausible range $0 \le w_{re} \le 0.25$.

In case of K-inflation with power-law kinetic term (\ref{lagnc}) the bounds on $N_k$ and $r$ for 
quadratic and quartic potential are shown in Table.~\ref{tab:rnsnkpophitwo} and 
Table~\ref{tab:rnsnkpophifour} respectively. 
We find that,  with $\alpha=4$ for quadratic potential, the tensor-to-scalar 
ratio $r>0.0740$ for $w_{re} \le 0.25$ and $r>0.0610$ for $w_{re} \le 1$, 
which is slightly greater than  the joint BICEP2/Keck array and  Planck-2018 
bound $r<0.06$ \cite{Ade:2018gkx}. However, this potential is compatible
with Planck-2018 bounds on $r$ for physically plausible range $0 \le w_{re} \le 0.25$ with 
larger values of $\alpha$. The $r-n_s$ predictions for these potentials are 
shown in Fig.~\ref{fig:rnsphitwoandfour}.  
It can be seen from the figure that, for these predictions to lie within
Planck-2018 $1\sigma$ constants,  the reheating equation of state $w_{re}\ge 0.25$ for quartic
potential. The bounds on $N_k$ and $r$ for exponential potential with power-law kinetic term
are shown in Table.~\ref{table:rnsnkexp}. We find that, for $\alpha=4$, the tensor-to-scalar ratio 
$0.1395\ge r \ge 0.1204$ for physically plausible range $0 \le w_{re} \le 0.25$ and 
$r\ge 0.0999$ for $w_{re} \le 1$, which are quite larger than the joint
BICEP2/Keck array and Planck-2018 bounds $r<0.06$ \cite{Ade:2018gkx}. Again bounds 
on $r$ are compatible with Planck-2018 bounds for larger values of $\alpha$. 
The $r-n_s$ predictions for exponential potential are shown in Fig~\ref{fig:rnsexp}. 
It is evident from the figure that, for these predictions to lie within joint 68\% constraints from Planck-2018 
observations, the effective equation of state during reheating should be greater than $1$.

These models of K-inflation are well motivated from string theory, 
and they  have similar  $r-n_s$ predictions. By imposing constraints from reheating
we can remove this degeneracy. In \cite{Sahni:1990tx,Mishra:2021wkm} it is shown that the spectrum of gravitational waves
generated during inflation is sensitive to the equation of state during reheating. We find
different allowed values of $w_{re}$ for different models to satisfy joint $68\%$ and $95\%$
C.L. constraints on $r-n_s$ from Planck-2018 observations. Hence, our analysis with 
future detection of gravitational waves can help us to find suitable model of inflation with 
noncanonical kinetic term.

\end{document}